\documentclass[12pt]{article}
\usepackage{latexsym}
\usepackage{amssymb}
\usepackage{graphicx}

\def\be{\begin{equation}}
\def\ee{\end{equation}}
\def\bear{\begin{eqnarray}}
\def\eear{\end{eqnarray}}
\def\nn{\nonumber}

\newcommand\bra[1]{{\langle {#1}|}}
\newcommand\ket[1]{{|{#1}\rangle}}

 \def\IZ{\relax\ifmmode\mathchoice
 {\hbox{\cmss Z\kern-.4em Z}}{\hbox{\cmss Z\kern-.4em Z}}
 {\lower.9pt\hbox{\cmsss Z\kern-.4em Z}}
 {\lower1.2pt\hbox{\cmsss Z\kern-.4em Z}}\else{\cmss Z\kern-.4em Z}\fi}
 \def\IB{\relax{\rm I\kern-.18em B}}
 \def\IC{{\relax\hbox{$\inbar\kern-.3em{\rm C}$}}}
 \def\Ic{{\relax\hbox{$\inbar\kern-.22em{\rm c}$}}}
 \def\ID{\relax{\rm I\kern-.18em D}}
 \def\IE{\relax{\rm I\kern-.18em E}}
 \def\IF{\relax{\rm I\kern-.18em F}}
 \def\IG{\relax\hbox{$\inbar\kern-.3em{\rm G}$}}
 \def\IGa{\relax\hbox{${\rm I}\kern-.18em\Gamma$}}
 \def\IH{\relax{\rm I\kern-.18em H}}
 \def\II{\relax{\rm I\kern-.18em I}}
 \def\IK{\relax{\rm I\kern-.18em K}}
 \def\IP{\relax{\rm I\kern-.18em P}}

 \font\cmss=cmss10 \font\cmsss=cmss10 at 7pt
 \def\IR{\relax{\rm I\kern-.18em R}}

\def\bra{\langle}
\def\ket{\rangle}

\def\D{\Delta}

\def\f{\phi}
\def\F{\Phi}

\newcommand{\sm}[1]{\mbox{\scriptsize #1}}
\newcommand{\tn}[1]{\mbox{\tiny #1}}
\renewcommand{\@}[1]{\sqrt{#1}}
\renewcommand{\le}[1]{\label{#1}\end{eqnarray}}
\newcommand{\bea}{\begin{eqnarray}}
\newcommand{\eea}{\end{eqnarray}}

\newcommand{\eq}[1]{(\ref{#1})}
\def\nn{\nonumber\\}

\def\ffract#1#2{\raise .35 em\hbox{$\scriptstyle#1$}\kern-.25em/
\kern-.2em\lower .22 em \hbox{$\scriptstyle#2$}}

\def\half{{1\over2}\,}
\newdimen\tableauside\tableauside=1.0ex
\newdimen\tableaurule\tableaurule=0.4pt
\newdimen\tableaustep
\def\phantomhrule#1{\hbox{\vbox to0pt{\hrule height\tableaurule width#1\vss}}}
\def\phantomvrule#1{\vbox{\hbox to0pt{\vrule width\tableaurule height#1\hss}}}
\def\sqr{\vbox{%
  \phantomhrule\tableaustep
  \hbox{\phantomvrule\tableaustep\kern\tableaustep\phantomvrule\tableaustep}%
  \hbox{\vbox{\phantomhrule\tableauside}\kern-\tableaurule}}}
\def\squares#1{\hbox{\count0=#1\noindent\loop\sqr
  \advance\count0 by-1 \ifnum\count0>0\repeat}}
\def\tableau#1{\vcenter{\offinterlineskip
  \tableaustep=\tableauside\advance\tableaustep by-\tableaurule
  \kern\normallineskip\hbox
    {\kern\normallineskip\vbox
      {\gettableau#1 0 }%
     \kern\normallineskip\kern\tableaurule}%
  \kern\normallineskip\kern\tableaurule}}
\def\gettableau#1 {\ifnum#1=0\let\next=\null\else
  \squares{#1}\let\next=\gettableau\fi\next}

\begin{document}
\pagestyle{empty}

\centerline{{\Large \bf Dualities and Emergent Gravity:}}
\vskip .1truecm
\centerline{{\Large\bf Gauge/Gravity Duality}}

\vskip 1truecm

\begin{center}
{\large Sebastian de Haro}\\
\vskip 3truemm 
{\it 
Department of History and Philosophy of Science\\ University of Cambridge\\
Free School Lane, Cambridge CB2 3RH, United Kingdom\\
{\tt{sd696@cam.ac.uk}}\\
and\\
Amsterdam University College\\
University of Amsterdam\\
Science Park 113, 1090 GD Amsterdam, Netherlands\\
{\tt{s.deharo@uva.nl}}\\
}

\end{center}
\vskip 0.5truecm

\begin{center}

\textbf{\large \bf Abstract}
\end{center}

In this paper I develop a framework for relating dualities and emergence: two notions that are close to each other but also {\it exclude} one another. I adopt the conception of duality as `isomorphism', from the physics literature, cashing it out in terms of three conditions. These three conditions prompt two conceptually different ways in which a duality can be modified to make room for emergence; and I argue that this exhausts the possibilities for combining dualities and emergence (via coarse-graining).

I apply this framework to gauge/gravity dualities, considering in detail three examples: AdS/CFT, Verlinde's scheme, and black holes. 
My main point about gauge/gravity dualities is that the theories involved, {\it qua} theories of  gravity, must be background-independent. I distinguish two senses of background-independence: (i) minimalistic and (ii) extended. I argue that the former is sufficiently strong to allow for a consistent theory of quantum gravity; and that AdS/CFT is background-independent on this account; while Verlinde's scheme best fits the extended sense of background-independence. I argue that this extended sense should be applied with some caution: on pain of throwing the baby (general relativity) out with the bath-water (extended background-independence). Nevertheless, it is an interesting and potentially fruitful heuristic principle for quantum gravity theory construction. It suggests some directions for possible generalisations of gauge/gravity dualities.

The interpretation of dualities is discussed; and the so-called `internal' vs.~`external' viewpoints are articulated in terms of: (i) epistemic and metaphysical commitments; (ii) parts vs.~wholes.

I then analyse the emergence of gravity in gauge/gravity dualities in terms of the two available conceptualisations of emergence; and I show how emergence in AdS/CFT and in Verlinde's scenario differ from each other. Finally, I give a novel derivation of the Bekenstein-Hawking black hole entropy formula based on Verlinde's scheme; the derivation sheds light on several aspects of Verlinde's scheme and how it compares to Bekenstein's original calculation.

\newpage
\pagestyle{plain}

\tableofcontents
\newpage

\section{Introduction}

\subsection{Two views on the emergence of gravity}\label{twov}

Recent developments in string theory are deeply transforming the way we think about gravity. In the traditional unification programme, gravity was a force which was to be treated on a par with the other forces: the aim was for a unified description of the four forces; and strings seemed to be of help because different vibration modes of the string give rise to different particles. But with the advent of holographic ideas\footnote{See 't Hooft (1993) and Maldacena (1998).}, a slightly different view is emerging that seems to be both more concrete and more modest in its approach. The new view starts from the realisation that gravity may be special after all, admitting a holographic reformulation that is generally not available in the absence of gravity: a gauge-theoretic reformulation in one fewer dimension; hence the name `gauge/gravity duality'. Thus the general goal of understanding gravity at high energies is now,
%added (alpha)
in the context of string theory,
%added
best conceptualised as consisting of two steps: 1) reformulate gravity (holographically) in terms of other forces, specifically: quantum field theories; 2) extract from this reformulation how gravity can be quantised. Progress on the first step over the past seventeen years has been impressive; whether the second step is actually needed is still a matter of debate. 

It is the latter question, on the necessity to quantise gravity, that prompts the main topic of this paper: the relation between dualities and emergence; and whether gauge/gravity dualities satisfy the usual standards for theories of `quantum gravity'. Broadly speaking, when there is a holographic duality---when gravity is dual to a quantum field theory---it makes sense to ask how to `reconstruct' the bulk,\footnote{de Haro et al.~(2001).} including the quantum manifestations of the gravitational force. 

And on this question, there are two main views: first, if the bulk and the boundary are exactly dual to one another, then the duality map can be used to find out what string theory or quantum gravity look like in the bulk. Gravity may be emergent or not; but the presence of duality up to arbitrarily high energy scales guarantees that it makes sense to speak of a theory of `quantum gravity'. 

The second, contrasting, view claims that gravity `emerges' from the boundary, without being exactly dual to it. In this case, gravity and the bulk are the product of an {\it approximate} reconstruction procedure.\footnote{This is also the guiding idea in other approaches to quantum gravity, such as analogue models of gravity and group field theory, which do not attempt to quantize the gravitational field itself; but rather regard it as emergent from some underlying non-gravitational fine-grained structure.} On this view, gravity does not exist at the fine-grained level, but exists only by grace of its emergence from the fine-grained degrees of freedom. Hence the guiding idea here is not duality but coarse-graining. On this view, step 2) above is superfluous; indeed, nonsensical: all there is to gravity is its reformulation in terms of some deeper (non-gravitational) theory; but asking how to quantise gravity is as futile (or as useful) as asking how to quantise water waves. Of course, the claim does come with a prediction: namely, that fundamental gravitons (or fundamental closed strings, for that matter) will never be found.

It is not the purpose of this paper to assess the relative merits of these contrasting viewpoints on the quantum gravity programme.\footnote{One reason I will not do this is because a final answer to this question depends on the answer to physical questions that are not yet settled.} Instead, the aim is to analyse them separately, thereby contrasting the different roles and conceptualisations that these approaches assign to `duality' and `emergence' of gravity and space-time. Regarding duality, I will concentrate on gauge/gravity duality (the duality between a gauge theory and a gravity theory) and its best-studied realization, the so-called AdS/CFT correspondence (section \ref{adscftc}). Some salient philosophical consequences of duality for AdS/CFT were already explicated in a previous paper, herein after referred to as Dieks et al.~(2015). In that paper it was pointed out how, borrowing ideas from the `emergence' camp, gravity can be seen to emerge in AdS/CFT as well; this requires an additional ingredient, which was identified as coarse graining. 

Accordingly, in the current paper, I will investigate in more detail the relation between dualities and emergence; for a priori, the presence of a duality {\it precludes} the phenomenon of emergence (the argument is presented in \ref{fall}). From my articulation of the notion of duality into three conditions (in section \ref{adscftc}) it will follow that emergence, realised when there is {\it duality broken by coarse-graining}, can only take place in two ways. I will use this simple framework for dualities and emergence to exhibit the mechanism for emergence in three important examples: AdS/CFT, Verlinde's scenario, and black holes. 

The application of dualities to AdS/CFT in section \ref{adscftc} will lead to the study of one of the requirements that are usually imposed on theories of quantum gravity: background-independence (sections \ref{bi}-\ref{extended}); and which here is required for AdS/CFT to qualify as a `gravity/gauge duality'. There will be two accounts of background-independence: (i) a minimalist one and (ii) an extended one. I will argue that only the first is strictly necessary for a theory of gravity (modelled after general relativity). AdS/CFT will be seen to be background-independent on this account. Verlinde's scheme, on the other hand, seems to suggest a background-independence more akin to the extended notion (section \ref{biV}). 

I will discuss the interpretation of dualities (section \ref{interp}) and articulate a choice of `internal' vs.~`external' viewpoints (Dieks et al.~(2015)) in terms of two relevant factors: (i) epistemic and metaphysical commitments; (ii) parts vs.~wholes. I will identify a case in which {\it only} the internal viewpoint is available.

I will develop the framework of emergence for approximate dualities in sections \ref{fall}-\ref{emapprox} and apply it in three examples of gauge/gravity relations: AdS/CFT, black holes, and Verlinde's scheme. I will make some of the underlying assumptions of Verlinde's scheme explicit (section \ref{Vscheme}) and provide a new derivation (section \ref{BH}) of the Bekenstein-Hawking black hole entropy formula based on Verlinde's scheme. I will discuss its significance, compare it to Bekenstein's original calculation, and discuss the extent to which this is a clear case of emergence (section \ref{emgr}).

Before these two main jobs, I turn in the rest of this introduction to the philosophical motivation (section \ref{care}), and a summary of basic facts about AdS/CFT (section \ref{notation}).

\subsection{Should philosophers care?}\label{care}

Why should philosophers care about these specialised branches of theoretical physics in which the dust has not yet settled on several central theoretical questions? I think there are two main reasons.

First: my comments above clearly reflect different attitudes that physicists take in their approaches to the problem of quantum gravity: there are those who believe that gravity needs to be quantised; and those who believe gravity is an effective description of what in essence is a gravitation-free microscopic world. In the absence of any experiments, these assumptions are necessarily theoretical and ontological; they involve a stance about whether gravity exists at the fine-grained level or whether gravity only emerges as a coarse-grained phenomenon. 
%added
Thus philosophical analysis is of help here.
%And philosophers can help analyse those presuppositions and make sense of the different notions of `emergence', `duality', etc., that physicists are using. So philosophy can be helpful as a tool for physicists to sharpen their arguments and to develop heuristic schemes from which to build new theories; and occasionally philosophy can also help identify some misguided assumptions in the heuristics, as we will see in the next subsection.

The second reason I believe philosophers should care about gauge/gravity duality is because there are interesting philosophical questions here: about duality; and about emergence of space, time, and gravity. Whatever the answers to the main questions about AdS/CFT---whether the duality breaks down due to some non-perturbative effect or not; or whatever answers about the ultimate nature of the objects that populate the bulk we may one day get---the result that the two sides of the duality are related by a map such as the one presented in $\tt{[AdSCFT]}$ (section \ref{notation} below) is robust, i.e.~well established theoretically. And this result also significantly impacts upon concepts, such as `duality' and `emergence' of gravity, that philosophers of physics have thought about. So it is worth taking examples such as AdS/CFT and Verlinde's scheme as case studies; so that general philosophical concepts can be analysed and set to work in concrete physical applications.
%The insight\footnote{Dieks et al.~(2015), sections 3.2.2 and 4.3.} that the duality of gravity does not by itself entail its emergence, and the analysis of the conditions under which emergence can take place, should by itself be indeed interesting. I believe there are other such notions that deserve philosophical scrutiny, such as background-independence. 

The paper is aimed at philosophers of physics but it is not meant as a self-contained introduction to gauge/gravity duality: although in section \ref{notation} I summarize the main facts that I will use later. For an introduction to this topic, as well as other background on which this paper is based, I refer the reader to Dieks et al.~(2015); other interesting philosophical work on AdS/CFT is Rickles (2011, 2012). There are also many excellent reviews of AdS/CFT: see for instance Ammon et al.~(2015). 

\subsection{Brief introduction of the gauge/gravity dictionary}\label{notation}

To fix ideas, and for future reference, I will now summarize some facts about gauge/gravity relations:\\
\\
$\bullet$ The AdS/CFT correspondence is a duality (a one-to-one mapping between states and quantities of two theories: see section \ref{cond}) between (i) string theory in anti-de Sitter space-time (AdS): the maximally symmetric solution of Einstein's equations with a negative cosmological constant; and (ii) a conformal field theory (CFT), i.e.~roughly: a quantum field theory with scale invariance, on the conformal boundary of this space. Hence the names `bulk' and `boundary', for the space-times where the theories are defined, respectively.\\
\\
$\bullet$ Generically, neither is the bulk pure AdS, nor is the boundary theory exactly conformal. It is sufficient that the bulk be asymptotically locally AdS and that the QFT on the boundary have a fixed point. Hence the more general name `gauge/gravity duality', though I will use the common phrase `AdS/CFT correspondence' when referring to specific models. Other examples of gauge/gravity relations that I will discuss, include: Verlinde's scenario and five-dimensional black holes. The latter are approximately dual to the physical conditions created at the RHIC experiment in Brookhaven, NY, and described by QCD. For these examples, cf.~respectively sections \ref{emapprox}, \ref{emgr}.\\
\\
$\bullet$ Fields in the bulk (whether scalar, vector, or tensor) are generically denoted $\f(r,x)$, where $r$ is the radial coordinate and $x$ are coordinates along the boundary directions. The dimension of the boundary is $d$; the bulk has dimension $d+1$.\\
\\
$\bullet$ The radial coordinate $r$ is dual to the renormalization group (RG) scale (see sections \ref{Vscheme} and \ref{Vscheme}) in the boundary theory.\\
\\
$\bullet$ The conformal boundary of AdS is at $r=0$. A boundary condition for the bulk field $\f$ is of the type: $r^{\D-d}\,\f(x,r)|_{r=0}=\f_{(0)}(x)$, where $\f_{(0)}(x)$ is fixed.\\
\\
$\bullet$ In the boundary theory, $\f_{(0)}(x)$ is interpreted as a source that couples to an operator ${\cal O}(x)$. The scaling dimension of this operator is denoted $\D$.\\
\\
Given the above, the AdS/CFT correspondence now relates the string theory partition function in a space with a negative cosmological constant and given a certain choice of boundary conditions for the fields $\f(r,x)$; to the generating functional of connected correlation functions for the corresponding operator ${\cal O}(x)$ in the CFT, coupled to the source $\f_{(0)}(x)$ as follows:
\bea
\hspace{3cm}Z_{\sm{string}}\left(r^{\D-d}\f(x,r)\Big|_{r=0}=\f_{(0)}(x)\right)=\Big\bra e^{\int\sm{d}^dx\,\f_{(0)}(x)\,{\cal O}(x)}\Big\ket_{\sm{CFT}}~.\hspace{.8cm} \tt{[AdSCFT]}\nonumber
\eea
This formula summarises the AdS/CFT correspondence. In section \ref{adscftc}, after analysing dualities in general, I will be concerned with the question of how precise this mapping is, as well as with its interpretation.

\section{Conditions for Duality and AdS/CFT}\label{adscftc}

In this section I discuss a notion of duality that captures its use in the physics literature, as an isomorphism: a one-to-one mapping between two theories, preserving certain structures of states and quantities (see de Haro et al.~(2015); Dieks et al.~(2015)). One can say that a duality is an equivalence between two theories, which agree on the values (in corresponding states) of all physical quantities, but may otherwise have very different formulations. Dualities thus have two main properties: one regarding the objects being related (two theories) and the other regarding the relationship between those objects (and their structures). I will cash out these two properties as three disconnected conditions to be satisfied by a duality: numerical completeness (of states and quantities), consistency (of the theories), and identity (of the invariant physical quantities). These conditions can be used to assess whether in a given case there is in fact a duality, which I will do (in later sections, \ref{Vscheme} et.~seq.) in three examples: AdS/CFT, applications of gauge/gravity duality to black holes, and Verlinde's scenario. Here, in section \ref{adscftc}, I will discuss completeness and consistency of CFT's in section \ref{completeness}; and completeness and consistency of the bulk theory in section \ref{bulkc}. There will be some surprises here, in that AdS/CFT, when the number of boundary dimensions is even, is background-independent in one sense but not in another. But this holds true for different reasons than are normally given in the literature: and the sense, in which AdS/CFT is background-independent, is sufficient to establish its consistency as a gauge/gravity duality. 

 In section \ref{interp} I will discuss how the interpretative fork articulated in Dieks et al.~(2015) (called: external vs.~internal viewpoints) relates to: (i) a physical difference about the kind of world theories describe; (ii) epistemic and metaphysical positions presupposed by the interpretation. I will spell out one case in which the internal view is the only one available. The concepts developed in this section will be used in section \ref{emergent} to develop a framework for the emergence of gravity.

\subsection{Conditions for duality}\label{cond}

I will now unpack the notion of duality described at the beginning of this section, so as to see the reasons that could lead to a gauge/gravity relation failing to be a duality; but keeping the conception sufficiently general that it can be applied to any duality. Two conditions must be met in order for the relevant isomorphism to exist. The laws of the two theories, and their (sub-) structures of quantities, should satisfy the following requirements:
\begin{enumerate}
\item {\bf Complete and consistent}. `Completeness' of the dynamical laws here means that they can be used in any relevant physical situation; `consistency' is construed as absence of internal contradictions (e.g.~breaking of some fundamental symmetry principle). With regard to the quantities: no quantities can be written down other than those that figure in the duality. In gauge/gravity dualities, this concerns the quantities mentioned in $\tt{[AdSCFT]}$ (end of section \ref{notation}): this sub-structure of quantities, evaluated on all states, contains what the two theories, on each side of the duality, regard as all the invariant `physical quantities'.\footnote{`Invariant' indicates that some quantities that are frame-dependent may nevertheless be physical. However, these can normally,  in one way or another, be related to the quantities in $\tt{[AdSCFT]}$. The distinction will therefore not concern us here.}
%added
Thus I spell  this out further in terms of the following two interdependent conditions on the states and quantities:\\
{\bf (Num)} Numerically complete: the states and quantities related by the duality (e.g.~those in $\tt{[AdSCFT]}$) are {\it all} the `relevant' quantities and states of the theory, independently on each side. `Relevance' is construed as:\\
{\bf (Consistent)} The dynamical laws and the quantities of each theory (leading to the formulation of e.g.~$\tt{[AdSCFT]}$) satisfy all the physical requirements of these theories, independently on each side. On the gravity side, the pertinent consistency condition is background-independence (see section \ref{bi}). On the CFT side, the quantities should have the correct properties under conformal rescalings (namely that they form a representation of the conformal group, up to an anomaly).
\item {\bf (Identical)}: the (sub-) structures of the invariant physical quantities on either side are identical to each other. In other words, the duality is {\it exact}.
\end{enumerate}

Condition (ii) regulates the {\it relationship} between the two sides.
If condition (ii) is not met, we have a weaker form of the correspondence: a relation that is non-exact (this was discussed in Dieks et al.~(2015)); there may be only {\it approximate} duality.\footnote{See Aharony et al.~(2000) p. 60 for a discussion of weak forms of AdS/CFT.} This would be the case if, for instance, the physical quantities  agree with each other in some particular regime of the coupling constants, but not beyond.\footnote{From the point of view of the {\it isomorphism} between the theories, (Identical) postulates the existence of an exact, structure-preserving map; whereas (Num) ensures that the domain and codomain of this map are identical with the physical content of the theories in question (states and quantities). See de Haro et al.~(2015).}

Condition (i) is a property that each side of the duality must satisfy independently: namely, there are complete and consistent theories on each side. The condition (Consistent) is relative to an appropriate body of phenomena; being (Consistent) thus differs from being the {\it final} theory of the world (whatever that might be), not even the final theory of a relevant class of phenomena. Theories can have limited applicability, yet satisfy all physical requirements relevant to the phenomena in question. For instance, Newton's theory of gravity is limited in at least two ways: (a) it is restricted to a regime of low speeds; (b) it makes nonsensical predictions (namely, of infinitely strong forces) when the distance between two masses goes to zero. The class of phenomena it describes is thus restricted to, at least: (a') low speeds compared to the speed of light; (b') large distances compared to (roughly) the Planck length. But relative to such phenomena, domain of parameters, and mathematical approximations made, it is a (Consistent) theory. This will motivate, in section \ref{emergent}, one of the two ways in which emergence can take place in a duality.

Thus, we have to specify which is the class of phenomena relevant to gauge/gravity dualities. I will spell this out in sections \ref{completeness} and \ref{bulkc}. Let me here mention that the relevant class of phenomena for AdS/CFT  is a tall order, as already evident from Maldacena's seminal work: for (Consistent) entails showing that on the bulk side there is a consistent theory of quantum gravity\footnote{A string theory, or M-theory, depending on the number of dimensions.} (but again, not the {\it final} theory). Such theories are expected to be background-independent. In sections \ref{bi}-\ref{diff} I will analyse what we mean by background-independence; and give a perturbative argument to the effect that AdS/CFT is as background-independent as one would wish it to be, in a suitably minimalist sense. 
I will also analyse a second reason for which duality could  fail---the failure of the relation $\tt{[AdSCFT]}$ to capture {\it all} the quantities in one theory (or in both of them), i.e.~a failure of (Num). I will address (Num) and, for the boundary side, (Consistent), in the the next subsection.

\subsection{Boundary theory: completeness and consistency of CFT's}\label{completeness}

In this subsection I discuss the first desideratum for a bijection---(Num)-(Consistent), independently on each side; leaving the (Consistent) {\it bulk} condition for the next subsection. 

%added$
On the CFT side, conditions (Num)-(Consistent) are fairly uncontroversial; the only possible threat coming from the generic absence of the concept of {\it free states}, in interacting CFT's, and the related impossibility of defining an $S$-matrix. If the $S$-matrix were a relevant physical quantity of any quantum field theory, the lack of it in a CFT would be a threat to (Consistent); as well as a breach of (Num). As I argue below, it is neither.

Let me first explain why, generically, there is no $S$-matrix in an interacting CFT. The construction of an $S$-matrix is based on the notion of `asymptotic states', i.e.~roughly: quantum states of particles that are far apart from each other and hence can be treated as non-interacting. `Can be treated as' here means that states of the full theory converge sufficiently rapidly\footnote{Haag's asymptotic theorem gives a measure for strong convergence of the time-dependent state vectors to the in/out states constructed from free states. See e.g.~Duncan (2012), section 9.3. Unfortunately, Haag's theorem only holds for a very restricted class of theories: those that do not have a mass gap.} to free states for large past or future times; at which times the particles are also sufficiently far away from each other. The measure that establishes such convergence is given in terms of the relevant length/mass scales in the theory: for instance, the mass of the particle excitation under consideration. In a CFT, however, there is no mass and no standard for when two particles are `far apart': conformal invariance implies the lack of a length scale in the theory to compare to. Any two points separated by arbitrarily `large' distances can always be brought arbitrarily `close' to each other by a conformal transformation. The mechanism by which interactions are `turned off' (in the sense just described) in ordinary quantum field theory, is therefore not available: a field excitation can never be fully isolated as a free particle. Scale invariance thus makes interacting CFT's very different from ordinary field theories: they do not possess the most central physical quantity.

There is, however, a well-defined notion of physical quantities in CFT's: the set of correlation functions of all local, self-adjoint, renormalizable operators that can be constructed and are appropriately invariant under all the symmetries of the theory.  Such operators correspond to the notion of an `observable', or `quantity', in the quantum mechanical sense, properly applied to a quantum field theory. And according to the AdS/CFT correspondence (cf~section \ref{notation}), the entire set of such correlation functions can be obtained by functional differentiation of $\tt{[AdSCFT]}$. 

Why is the lack of an $S$-matrix not a threat to (Num)-(Consistent)? Some field theorists have argued that the lack of an $S$-matrix is problematic from a physical perspective about what good physical quantities are supposed to be like.\footnote{Such criticisms were made by Gerard 't Hooft at the Spinoza Meeting on Black Holes (discussion with Juan Maldacena, 1999).} The lack of a notion of `free particle', the argument goes, is a threat to the ability of these theories to give good descriptions of nature. Free particles are not only crucial in formulating quantum mechanics and quantum field theories: the lack of them also constrains the kinds of experimental set-ups that can be achieved: if the world were described by a CFT, the experimenter would be unable to create the conditions of isolation and control of the system that are crucial in experiments such as the ones carried out at CERN. These conditions are also instrumental in ensuring the reproducibility of measurements. 

This objection, however, seems to relate to the {\it pragmatics} of describing our actual world, rather than to a failure of (Consistent) in the structure of conformal field theories. Important as such intuitions may be as heuristic guiding principles: criticism based on them does not seem to address the CFT's themselves, or indeed AdS/CFT; neither of which are advanced as actual descriptions of the world---where, after all, the cosmological constant seems to be positive rather than negative---but as models, in the case of AdS/CFT, for relating quantum gravity theories and quantum field theories.\footnote{An important criticism advanced by 't Hooft is that string theory (including some of the CFT's involved) has not been given a formulation with a mathematical rigour comparable to that of the standard model, where at least lattice approaches are available ('t Hooft (2013)).} The question that is relevant for an assessment of a CFT by its own lights is, therefore, not whether the physical quantities satisfy standards of current scientific practice, but whether it could give a candidate physical representation of a world where the fields satisfy classical conformal invariance (alternately, using AdS/CFT, whether it could give a candidate physical representation of a world with a {\it negative} cosmological constant). Thus, by the CFT's own standards, (Num)-(Consistent) are satisfied. 

%added
Furthermore, the duality shows the worry to be misplaced---with respect to the boundary theory there are no free particles; but that same physics does correspond to local physics in the bulk. 

\subsection{Bulk theory: completeness and consistency}\label{bulkc}

Next I turn to the {\it bulk} theory and whether it satisfies (Num)-(Consistent). In \ref{numbulk} I discuss its (Num)-erical completeness; and in \ref{bi}-\ref{diff} the (Consistent) condition.

\subsubsection{(Num) in the bulk theory}\label{numbulk}

There have been suggestions in the literature that (Num)-erical completeness of the bulk theory requires the definition of non-local quantities. In particular, $\tt{[AdSCFT]}$ (end of section \ref{notation}) only envisages quantities that are defined near the boundary, whereas proposals such as Heemskerk and Polchinski (2011) suggest that operators need to be defined in the interior of AdS as well. If so, equation $\tt{[AdSCFT]}$ would be incomplete. Whether (Num) could still be met, if such proposals are correct and necessary, would thus depend on whether there is an equation that generalizes $\tt{[AdSCFT]}$ and relates the new quantities on both sides of the duality. The proposal of Heemskerk and Polchinski (2011) is certainly in this spirit: it is assumed that $\tt{[AdSCFT]}$ can be {\it extended} to include such operators, hence preserving (Num) of the augmented theories.

\subsubsection{(Consistent)-laws: Background-independence of the laws}\label{bi}

%added-rephrased
I now discuss how the bulk side of the duality satisfies (Consistent). The main physical constraint to be imposed on theories of quantum gravity is that they be `background-independent'.\footnote{There is no claim here that background-independence exhausts the desiderata (Consistent), on the bulk side, in AdS/CFT. I am claiming that it is an important desideratum that has not been analysed with sufficient care, in the literature, so far.} This expectation stems from Einstein's theory of general relativity: where the metric is a dynamical quantity, determined by the equations of motion, rather than set to a fixed value from the outset; and this, in particular, implies the familiar statement that there is no preferred class of reference frames. (Thus, background-independence has been taken to be a requirement that summarises several of the desiderata of theories of quantum gravity: covariance, lack of a fixed geometry, etc.)\footnote{Unfortunately, the concept of `background-independence' is not a precise one with a fixed meaning. For a discussion of this, see Belot (2011). According to Belot, the concept is relative to an interpretation of the theory. Between a background-dependent theory and a completely background-independent theory there is a range of possibilities: background-independence comes in degrees. See also Giulini (2007). For a seminal discussion of background-independence, see Anderson (1964) and (1967).\label{refbi}} I endorse this consensus, for Einstein's theory is indeed our best guide to laws and quantities for gravitation that are (Consistent). But, while I adopt the consensus, I also argue that, for establishing (Consistent), it is important to distinguish:\\
\indent (i) A {\it minimalist sense} of background-independence, closely modelled on general relativity's own background-independence: which is sufficient to establish (Consistent) in gauge/gravity dualities. \\
\indent (ii) An {\it extended sense}, in which  the initial or boundary conditions are also to be dynamically determined. This sense, I will argue, is not necessary for the theory to be (Consistent), or indeed for duality to obtain.  

I will argue that one has to beware of promoting the extended sense to serve as an {\it a priori} standard for background-independence: for it would render general relativity background-dependent (see below): thus throwing the baby out with the bath-water. The extended sense nevertheless is a desirable {\it heuristic principle} towards constructing new theories of quantum gravity because it corresponds to practical needs. 

My warning is confirmed  by the philosophical literature: the concept of background-independence is not a precise one; in fact, there is vast disagreement about a preferred notion of background-independence (see footnote \ref{refbi}).

In the rest of this subsection I will focus on (i), leaving (ii) for section \ref{extended}. First I discuss the background-independence of the dynamical laws of the bulk theory; and how these are background-independent. Here, the main issue is that solving the equations requires asymptotic conditions. In section \ref{diff} I will discuss the much more subtle issue of the diffeomorphism invariance of the physical quantities.

The {\it minimalist sense} of---or approach to---background-independence (in first approximation: a refinement will follow shortly) is as follows. A theory is background-independent if: (a)  it is covariant; (b) all the relevant covariant fields (in particular, the metric) are determined dynamically from the equations of motion. This minimalist notion is sufficient for the current discussion of (Consistent) because a stronger sense would not be compatible with Einstein's equations with a negative cosmological constant. Indeed: in general relativity, with any sign of the cosmological constant, the theory generically does {\it not} determine its boundary or initial conditions; and thus there is a rich space of solutions. To fix a solution, boundary or initial conditions need to be stipulated: initial Cauchy data, in the case of zero cosmological constant; asymptotic boundary conditions, for negative values of the cosmological constant; and past- or future-timelike infinity data, for positive values.\footnote{Despite the significantly different global properties of the three cases, boundary or initial conditions for the metric are needed for {\it any} value of the cosmological constant. The need for asymptotic boundary conditions for quantum fields in AdS was discussed in the seminal paper Avis et al.~(1978). For boundary conditions in the conjectured dS/CFT (positive cosmological constant), see de Haro et al.~(2014) and references therein.}

In the minimalist sense, then, the bulk theory is fully background-independent: at the classical level, its dynamical laws, i.e.~its equations of motion, are Einstein's equations of general relativity with a negative cosmological constant. They can of course be derived from a generally covariant action, that is required to be invariant under diffeomorphisms that leave the asymptotic form of the line element fixed (notice the qualification: which will be crucial in sections \ref{diff}-\ref{extended}!). Adding (string-theoretical) classical matter fields does not change this picture: since these contribute covariant terms to the equations of motion.

What about quantum corrections? In the full quantum version of AdS/CFT, quantum corrections manifest themselves as higher-order corrections in powers of the curvature to the classical action. These are generally covariant as well. Hence, in this minimalist sense, also a {\it quantum} version of AdS/CFT---which would include quantum corrections as an infinite series with increasing powers of the curvature---is background-independent. In other words, the dynamical laws are, in a perturbative sense, (Consistent).

Can the study of {\it particular solutions} of Einstein's equations introduce some background-{\it de}pendence? The equations of motion do not determine the boundary conditions, which need to be specified additionally.\footnote{This elementary point has subtleties regarding the relation between the bulk and boundary metrics in the application to AdS/CFT, which are discussed as an essential ingredient of the `dictionary' between the bulk and the boundary in de Haro et al. (2001).} Notice that this is {\it not} a restriction on the class of solutions considered; the equations of motion simply do not contain the information about the boundary conditions; and the latter have to be supplied in addition to them.\footnote{Thus, for small curvatures, the configuration space considered by AdS/CFT is exactly the same as that considered by general relativity: setting matter fields to zero, the classical bulk theory {\it is} vacuum general relativity with a negative cosmological constant.} Indeed, as noted before, already the Einstein-Hilbert action in the presence of a boundary---which cannot be removed---is not invariant under arbitrary diffeomorphisms. As in classical mechanics, two initial conditions---the asymptotic values of the metric and of its conjugate momentum---are required in order to fully solve the equations of motion. 
%added
Yet it would be wrong to say that classical mechanics is dependent on a fixed choice of time coordinate, or is in some way not reparametrisation invariant, just because initial conditions are required in order to pick particular solutions of the equations of motion. It is clear that any theory whose laws are described by differential equations will require such choices of initial or boundary conditions. Thus, also in the presence of boundary conditions, the bulk theory is (Consistent).

I emphasize this point because it has been claimed  in the literature that AdS/CFT is not background-independent because of the boundary conditions.\footnote{See e.g.~Smolin (2005).} (The claims concern background-independence {\it tout court}: the notion is often not qualified: but that is problematic; for, as argued before, the claim dispenses with general relativity altogether). Another way to restate the above is to say that setting boundary conditions is {\it pragmatic}: the laws are universal but the boundary conditions vary from case to case.\footnote{Recently, Pooley (2015) has developed a helpful conception of background-independence (cf.~his section 8, version 3, with its emphasis on {\it dependent} variables), whose verdict on AdS/CFT is the same as mine: as being fully background-independent. His conception, however, does not seem to take into account quantum theories.} Which is not to say that boundary conditions are altogether philosophically irrelevant! For two specific reasons: (i) they play a key role in the duality, i.e.~the duality does not relate generic theories to each other; but theories with specific choices of boundary conditions, as seen from ${\tt[AdSCFT]}$ (boundary conditions in the bulk are related to sources on the boundary: fifth and sixth bullet points in section \ref{notation}); (ii) the interpretation of the duality, to be worked out in section \ref{interp}, depends on the meaning assigned to the boundary conditions.

Rather than viewing the dependence on boundary conditions as an instance of explicit breaking of background-independence, we should regard this as a case of {\it spontaneous breaking} of the symmetry: the symmetry is only broken by a choice of a particular solution because this entails a choice of asymptotic background metric. Further quantities one may wish to compute, such as $\tt{[AdSCFT]}$ and its derivatives, will carry some dependence on this choice of solution. These quantities, however, do exist (and match, in the semi-classical approximation between both sides of $\tt{[AdS/CFT]}$) for any background---even when the CFT is on a background that is not even locally Minkowski  (de Haro et al.~(2001)). The physical quantities mentioned in $\tt{[AdSCFT]}$ are scalars. In section \ref{diff} I will discuss the tensor quantities defined by functional differentiation of those in $\tt{[AdSCFT]}$.

Summing up: there are two reasons for not seeing boundary (or initial) conditions as backgrounds. (i) Unlike backgrounds, boundary conditions are {\it arbitrary}: they are not fixed but differ from solution to solution. They are not a priori elements of a theory; but are factors chosen pragmatically to describe a particular physical situation. (ii) They do not break the symmetries of theories; but rather parametrise the space of solutions: and are thus better identified as cases of {\it spontaneous} symmetry breaking, in the innocuous sense of `choosing a particular solution which happens to lack a certain symmetry possessed by the equations of motion'.

I have thus argued that: 1) the dynamical laws are background-independent; 2) the dependence on the boundary metric, introduced by the need to impose boundary conditions, is not a failure of background-independence in the minimalist sense, but merely a case of spontaneous symmetry breaking. Thus (Consistent) is established, by standard criteria that apply to general relativity.

\subsubsection{(Consistent)-quantities: Background-independence of the quantities}\label{diff}

In the previous subsection I argued that the dynamical laws and their boundary conditions are background-independent in a sufficiently strong minimal sense, hence  (Consistent). Now I move on to the quantities. Again, I stick to the minimalist sense needed for (Consistent), with its two aspects: (a) covariance; (b) the fields are dynamically determined from the equations of motion, up to boundary conditions: where the latter qualification has been added in light of the discussion in the previous subsection. 

Let me anticipate here that the case for (a) is subtle and will occupy the rest of this subsection. As for (b), within the aims I have set myself\footnote{That is, to produce a notion of background-independence that parallels that in Einstein's theory and is helpful in giving a verdict on the condition (Consistent) of the bulk theory.}: it is straightforward. The fields and physical quantities depend on the boundary conditions; but, as we saw, this is innocuous: it is not a breach of background-independence, nor consistency, that the quantities in a theory whose laws are given by differential equations, depend on the boundary conditions. 

Thus I turn to (a). Let me spell out, first of all, the relevant class of diffeomorphisms. The laws---Einstein's equations with perturbative corrections---are covariant under {\it any} bulk diffeomorphisms, as discussed in the previous subsection. But the {\it action} was not invariant under all diffeomorphisms. Neither are the physical quantities: being purely boundary quantities, they depend on boundary conditions. One must therefore distinguish:\footnote{See de Haro et al.~(2015), section 5.2.} (a1) diffeomorphisms that preserve the asymptotic form of the metric; (a2)  `large' diffeomorphisms---those that change the conformal class of the asymptotic metric. Notice that the diffeomorphisms (a2), though they are symmetries of the laws, are not---need not be---symmetries of the action and the physical quantities. Such diffeomorphisms relate physically {\it inequivalent} solutions. The only requirement here is that such diffeomorphisms map solutions to solutions. And that they do: for a solution of Einstein's equations is obtained, given any choice of boundary conditions for the argument solution.

Thus the relevant diffeomorphisms to be discussed are (a1), those that preserve the asymptotic form of the metric. They are the putative real symmetries of the theory: i.e.~not only symmetries of the laws, but also symmetries of the physical quantities. Restriction to such diffeomorphisms is in fact required by the derivation of the equations of motion from Einstein's action. So, to establish (Consistent), we have to see how such diffeomorphisms act on the physical quantities.

There are two basic types of physical quantities: scalars and tensors, all of them derived from the left-hand side of $\tt{[AdSCFT]}$. 
The bulk partition function itself, on the left-hand side of $\tt{[AdSCFT]}$, being a scalar, is diffeomorphism invariant for the diffeomorphisms (a1). Further  physical quantities are derived from $\tt{[AdSCFT]}$ by taking derivatives with respect to the boundary metric. These physical quantities transform as tensors under the diffeomorphisms discussed in the last pargraph. 

Here, however, there is a difference between the cases of even vs.~odd boundary dimension $d$:\\
\\
$\bullet$ For odd $d$, the physical quantities are covariant if they carry (boundary) space-time indices, and invariant if they don't.\footnote{Because it is only {\it boundary} quantities that are physical quantities, bulk scalars such as the Ricci scalar are {\it not} part of the physical quantities considered in $\tt{[AdSCFT]}$: as usual in background-independent theories.} This is precisely what one expects from tensor physical quantities in a generally covariant theory, hence diffeomorphism invariance is preserved by the physical quantities derived from $\tt{[AdSCFT]}$ in the case of odd $d$. \\
$\bullet$ For even $d$, the bulk diffeomorphisms that yield conformal transformations of the boundary metric are broken due to IR divergences. This is called the `holographic Weyl anomaly', or the `diffeomorphism anomaly', cf.~Henningson et al.~(1998). In the rest of this subsection, I will give some details (for more, cf.~Henningson et al.~(1998)). We will see that the holographic Weyl anomaly exactly mirrors the breaking of conformal invariance by quantum effects in the CFT---the well-known `conformal anomaly'.\\
\\
For even $d$, the asymptotic conformal invariance is broken by the regularization of the large-volume divergence. The partition function $\tt{[AdSCFT]}$ now depends on the conformal structure picked for regularization. As a consequence, physical quantities such as the stress-energy tensor derived from $\tt{[AdSCFT]}$, no longer transform covariantly, but pick up an anomalous term. Now, recall that the conformal anomaly is well-known in two-dimensional CFT's (see e.g.~sections 5.4.1-5.4.2 of Di Francesco et al., 1996), where the new term is the Schwarzian derivative. Such a term is also present  in four and six dimensions (Deser et al.~(1993)). The conformal anomaly of the CFT is a quantum effect, proportional to $\hbar$. On the gravitational side, the anomaly is inversely proportional to Newton's constant $G$ and is a classical effect resulting from the large-distance divergence of the action.  
%added
That a quantum effect can be mirrored by a classical effect on the dual side, is a consequence of the weak/strong nature of the duality (see Dieks et al.~(2015)): a perturbative, weak-coupling regime, in one theory, corresponds to a highly non-perturbative regime in the other. 
As with anomalies in field theory, there is no regularization of the boundary action that respects the asymptotic conformal symmetry.
Furthermore, the anomaly is robust: it is fully non-linear and can be derived without relying on classical approximations. 

In string theory the conformal anomaly needs to vanish in order for the theory to be consistent; and this is in fact how both the required number of space-time dimensions and Einstein's equations are derived in string theory.
%(for a discussion, see e.g.~Vistarini (2015): this volume). 
The reason the conformal anomaly needs to vanish there is that conformal invariance is part of the local reparametrization invariance of the world-sheet metric. The world-sheet metric is an auxiliary field that has been introduced, but it is integrated out and the theory cannot depend on it. This is not the case in a generic CFT. Such a theory is formulated covariantly, so that it can be used in any background; but the background is not selected dynamically from equations of motion. So, despite the covariance of the formulation, diffeomorphism symmetry is not part of the fundamental symmetry of the theory. A fortiori, there is no a priori reason why the conformal symmetry should remain at the quantum level.

Going back to the physical quantities in $\tt{[AdSCFT]}$: the covariance requirement (a) of the minimalist approach to background-independence, discussed at the beginning of this subsection, is preserved for odd boundary dimensions $d$, but violated for even dimensions by the holographic Weyl anomaly. The diffeomorphisms violated, however, are {\it not} true symmetries of the theory; they are boundary diffeomorphisms and, as discussed in section \ref{bi}, there is no reason to expect the CFT to be invariant under them.\footnote{This contradicts the claim, often seen in discussions of background-independence (and going back to Kretschmann's objection to Einstein's claims that general covariance selected general relativity), that `any' theory can be given a manifestly covariant formulation, given the right variables. This statement is {\it false} for quantum theories, as the conformal anomaly demonstrates; as well as for classical theories, such as general relativity, with boundaries.} Thus, for the same reasons discussed in the previous subsection: the physical quantities are background-independent in the minimalist sense.

%added
To summarize this subsection: among all the diffeomorphisms under which the laws are invariant, there are some (those that induce a  conformal transformation at infinity) that are broken by divergences associated with the infinite volume of the boundary, when the boundary dimension is even. The anomaly cannot be removed; and it does mean that the physical quantities are {\it not} invariant (covariant) under those diffeomorphisms. Yet this is {\it not} a breach of (Consistent): the quantities need not be invariant under {\it those} diffeomorphisms in the first place---an argument that is best made by comparison with the conformal anomaly in the CFT. Nor does the conformal anomaly lead, in the CFT's at hand, to any inconsistencies. Thus, background-independence of both dynamical laws and physical quantities, in the minimalist sense, is established.

Let me summarise the upshot of this subsection and the previous one, in the following two statements:\\
\\
$\bullet$ {\it CFTs are not background-independent for two reasons (depending on dimension): (i) the metric is not determined by the equations of motion; (ii) in an even number of dimensions, the theory is not covariant because of the conformal anomaly.}\\
\\
$\bullet$ {\it CFTs lack of background-independence does not contradict the background-independence of the bulk side of AdS/CFT: for the dependence is with respect to} different metrics.

\subsubsection{Minimalist vs.~extended senses of background-independence}\label{extended}

In section \ref{bi} I introduced the minimalist and extended senses of background-independence. The minimalist sense was constructed with the aim of assessing whether the bulk side of gauge/gravity dualities is (Consistent): where the physical requirement was that both the laws and the physical quantities should have the properties they have in general relativity. For that reason, this is the only notion of background-independence that one can {\it a priori} demand of a theory. 

It is to the extended notion of background-independence that I now turn. The question is: if the minimalist sense is geared towards being (Consistent) as in general relativity, could a stronger physical requirement be placed on (Consistent), thus producing a stronger version of background-independence: e.g.~for theories of quantum gravity? This is what I called the `extended sense' of the notion, and I will argue that it is not compelling, though it could be pragmatically helpful.

Let us first recall the meaning of the minimalist sense: (a) covariance; (b) the metric being determined by the equations of motion (up to boundary conditions). The latter addition, of allowing boundary (or initial!) conditions, was necessary because one is dealing with theories whose laws are differential equations (to be stipulated freely)---including general relativity, with any sign of the cosmological constant! 

This does indeed suggest how, in theories of {\it quantum} gravity, the idea of an {\it extended} sense of background-independence may be taken seriously as a heuristic guiding principle towards theory construction: namely, one might require that the boundary conditions should now be obtained dynamically rather than by stipulation. The Hartle-Hawking no-boundary proposal at the {\it beginning} of the universe is one such example. 

In fact, the extended sense of background-independence is not altogether alien to AdS/CFT, where much work has been done on generalising the boundary conditions from Dirichlet (the ones discussed here, with the metric fixed at infinity) to other boundary conditions, such as Neumann (where the conjugate momentum is held fixed instead, and the metric is allowed to fluctuate), mixed, and even more general cases (cf.~de Haro (2009) and references therein). This is certainly possible in odd boundary dimensions. In even boundary dimensions, the diffeomorphism anomaly seems to be a real threat here. 

Be that as it may, for such generalisations: they are not necessary for either: (i) gauge/gravity dualities to be genuine cases of quantum gravity theories; (ii) for duality to obtain. As a heuristic guiding principle, the extended sense can only be justified {\it pragmatically}: useful for its practical applications in quantum gravity calculations; but it should not be mistaken for an {\it a priori} requirement of any theory of quantum gravity: on pain of throwing the baby out with the bath-water.  I hope to come back to this issue elsewhere.

\subsection{Interpreting dualities}\label{interp}

Given the symmetry between the two ends of the duality relation---the one-to-one mapping of quantities and states allows us to replace each quantity on one side by a corresponding quantity on the other---is there a way to decide which theory is more adequate according to all the physical and epistemic criteria one should apply? Or does duality amount to full theoretical equivalence, so that we should speak of two formulations of one theory, rather than of two theories?

%added
The answer to all these questions cannot be given from the mathematical structure of the theories alone. For isomorphism, by itself, is not sufficient to establish full equivalence between the theories as descriptions of the world: for it does not give us an {\it interpretation} of the theory. The above questions can only be answered once a decision is made on several pertinent epistemic and metaphysical issues, as well as on the content of the physics described; in other words, once the duality is {\it interpreted} physically. In Dieks et al.~(2015) the basic interpretative fork was between external or internal points of view.\footnote{Related works that also discuss the interpretation of dualities, include: Matsubara (2013), Huggett (2015).} 

\subsubsection{Defending the internal point of view}\label{defend}

In this subsection I will start by restating these two points of view; and I will also: (a) discuss a number of epistemic and metaphysical issues underlying the distinction; (b) provide additional arguments in support of the internal view for the dualities under consideration; (c) discuss some examples. The framework that follows is thus not restricted to gauge/gravity dualities, but holds for any dualities satisfying the conditions (Num)-(Consistent)-(Identical)  formulated in section \ref{cond}. 

One can distinguish the following two viewpoints. They represent two mutually exclusive interpretative possibilities:\\
\\
(i) An {\bf external point of view,} in which the meaning of the physical quantities is externally fixed. For instance, in the context of AdS/CFT, the interpretation fixes the meaning of $r$ as the `radial distance' in the bulk theory, whereas in the boundary theory the meaning of $r$ (or whatever symbol it corresponds to in the notation of the boundary theory) is fixed to be `renormalization group scale', as reviewed in section \ref{notation}.\footnote{
%added
Strictly speaking, $r$ is not a diffeomorphism invariant quantity, and so it is not an physical quantity in the sense of $\tt{[AdSCFT]}$, explicated in (Consistent). It is, however, a physical quantity with respect to a choice of reference frame; and so good enough for the illustrative purposes of this example. In particular, the physical quantities in $\tt{[AdSCFT]}$ are evaluated making a choice of $r$, the end result being independent of this choice. Notice, however, that there is no question here of underdetermination the {\it values} of $r$: upon fixing a frame, the values are the same on both sides of the duality; it is only the physical interpretation (length vs.~energy scales) that differs, as discussed in the main text.}

More generally, on the external view the interpretative apparatus for the entire theory is fixed on each side. Since the two interpretations are different, the physics they describe are mutually exclusive. Hence only one of the two sides of the duality provides a correct description of a given set of empirical facts. On this interpretation there is only a formal/theoretical, but no empirical, equivalence between the two theories, as they clearly use different physical quantities; only one of them can adequately describe the relevant empirical observations. In short: the `exact symmetry' between the two theories, expressed by the duality relation, is broken by the different interpretation given to the symbols.\\
\\
(ii) An {\bf internal point of view:} if the meaning of the symbols is not fixed beforehand, then the two theories, related by the duality, can describe the same physical quantities. Indeed, on this view we would normally say that we have two formulations of one theory, not two theories. For example: for what might intuitively be interpreted as a `length', a reinterpretation in terms of `renormalization group scale' is now available. In other words, one remains undogmatic about the intuitive meaning of the symbols and derives their interpretation from their place and relation with other symbols in the theoretical structure.\footnote{This is reminiscent of the ``conceptual role semantics'' tradition in the philosophy of language.}

In general, the difference between the external and internal points of view seems to depend on two factors.

 The first are the epistemic and metaphysical assumptions under which one interprets the duality: for not all epistemic or metaphysical positions are supported by each of the viewpoints. Specifically, the internal viewpoint seems incompatible with a strong form of metaphysical realism in which one of two equivalent formulations describes the world better. The internal view is supported not only by anti-realism, but also by weak forms of realism, e.g.~structural realism (see Matsubara~(2013), section 3.2).\\

The second interpretational aspect is suggested from the wording `external' vs.~`internal'. `External' suggests that there is a relevant environment to which the theory may be coupled, an external context that fixes the meaning of the symbols. `Internal' suggests that such context is absent. Thus, it is tempting to see the difference between (i) and (ii) (i.e.~the difference in the way meaning is assigned to symbols) as matching a difference about the kind of theories considered:\\
\\
{\bf (Whole)}: theories of the whole world: i.e.~of the universe; and: \\
{\bf (Parts)}: theories that only provide partial descriptions of empirical reality: i.e.~that describe parts of the world. \\
\\
But, as I will now discuss, the link is less direct than it appears. I will maintain that the internal point of view is correct, assuming: (i) (Whole); and (ii) some form of structural realism (see the discussion in this subsection, below). If (i)-(ii) are met, it is impossible, in fact meaningless, to decide that one formulation of the theory is superior, since both theories are equally successful by  all epistemic criteria one should apply. And (Parts) indeed applies to ordinary cases of {\it underdetermination}: where two parts of the world are isomorphic yet clearly distinct.
 
But I also submit that the internal viewpoint is not restricted to (Whole): it may obtain in cases of (Parts). For in situations where two parts of the world are isomorphic yet distinct, there is usually a notion of the two parts  being `at different locations'; or of being instantiated in another such distinct way. For instance: under the relevant idealisations (small displacements, neglect of friction, etc.), the small angular swing of the clock pendulum in my neighbour's living room is isomorphic with the linear motion of the bob at the end of the spring that my colleague uses for her class demonstrations in classical mechanics. Yet the two systems are clearly distinct: one belongs to my neighbour, the other to my colleague. Location (only displacements were mentioned) or other features that make a difference to the two instantiations, are not part of the isomorphism: they are not captured by the descriptive idealisations; yet they are relevant aspects to the ontological distinction between {\it this} pendulum and {\it that} spring! (In gauge/gravity dualities, `location' and `space' are just some of the properties that do not have a fixed reference under the duality.) Thus, in so far as it can be argued that the elements of the world that the idealisation neglects are relevant to the ontology of the theory (e.g.~`location' of the pendulum or the spring), (Parts) will {\it not} support an internal viewpoint---the two situations are indeed physically distinct. But in so far as those elements can be argued to be irrelevant (within a certain context and-or approximation), (Parts) {\it will} support the internal viewpoint.

To illustrate this case, let me discuss a familiar example. Position-momentum duality in quantum mechanics, despite its differences with the dualities discussed here, may be a useful analogy.\footnote{For more on this duality as a case study, see Fraser (2015).} With the advent of quantum mechanics one might have insisted that a description of atoms in terms of position or in terms of momentum have different physical interpretations. In particular, the meanings of $p$ and of $x$ are different because the kinds of experimental set-ups that measure position and momentum are different. However, we have gotten used to the idea that the position and momentum descriptions really are different representations of the same theory---with Fourier transformation playing the role of the duality map. Given that all the physical quantities have the same values in both representations, we should simply say that we have not two theories but one.\footnote{Bohr's position apparently contradicts this because he invokes the context of the measurement as determining which concept---position or momentum---is applicable. On Bohr's view, for a given measurement context, there is only one possible description ($x$ {\it or} $p$, but not both: which one applies depends on the context). On such a view, however, the requirement (Consistent) seems to fail; and so this is not a case of duality but of {\it complementarity}. It would be interesting to discuss this distinction further. The comments here also do not apply to pilot wave theories, because their strong realist assumptions do not at all mesh with the internal viewpoint. As such, both examples reinforce my interpretative case.} Going back to our discussion; (a) position-momentum duality is indeed a duality according to the conditions in section \ref{cond}: notice that the $x$- and $p$-representations are both (Consistent) (they satisfy all physical requirements on physical quantities) and (Num) (all physical quantities can be described on each representation); by Fourier transformation, they are also (Identical). (b) Regarding the interpretation of the duality, quantum mechanics is clearly not an instance of (Whole), but (Parts). For instance, it does not contain gravity; hence it does not correctly describe some parts of the world, such as black holes, where gravity plays an important role. Yet we do not expect the incorporation of gravity to modify our conclusion about the equivalence of the position and momentum descriptions at the atomic scale, for those parts of the world to which quantum mechanics uncontroversially applies. So position-momentum duality in quantum mechanics, on a structuralist interpretation (in which `position' and `momentum' as such {\it can} be abstracted away from the fundamental ontology), appears to be a clear case of the internal view, despite the fact that quantum mechanics only provides a (Parts)-ial description of the world. And thus it is an example of the internal perspective on a (Parts)-ial theory. In fact, this duality is usually seen as teaching us something new about the nature of reality: namely, that atoms are neither particles, nor waves. By analogy, it is to be expected that gauge/gravity dualities teach us something about the nature of space-time and gravity.

The conditions (Whole) and (Parts) are also consistent with the theory giving an idealized description of the world,\footnote{The assumption that an idealization exists is in the duality condition (Consistent). Weaker conditions (of there being `no limiting system', or in which `the limit property and limiting system disagree') are in Bouatta and Butterfield~(2015).} rather than describing it in full detail: which is good, as most theories are like this! In other words, they describe a possible world similar to ours. Because, by assumption, this possible world is sufficiently close to ours (i.e.~the theory is (Consistent) with respect to the idealized world, and describes our world sufficiently well), the distinction (Whole)/(Parts) is preserved under the idealization. 

We see from this discussion that, in the framework for dualities introduced in section \ref{cond}, idealizations and the related approximations can be understood as shifts in the relevant body of phenomena stipulated in (Consistent): effectively replacing our world by a sufficiently close (ideal, or approximated) possible world within the condition (Consistent).

To summarize my interpretative case: given a duality, and given (Whole)\footnote{But not only! The internal point of view also applies to (Parts), as the example of quantum mechanics makes clear.} the internal point of view seems the correct one. It is also better suited to a `science first' position on metaphysics. 

\subsubsection{Pragmatics and broken dualities}\label{pragma}

The strong equivalence between the theories, that exists in the internal view, may be broken by the practical superiority of one theory over the other.
% These may be taken as a reason to resort to the external view. 
Pragmatic factors may make one formulation superior---for instance, if it allows certain computations to be easily done in a particular regime in one theory whereas they are hard to do in the other. This is certainly epistemically relevant and may in fact reveal some objective features of the underlying physical system, in the same way that the applicability of classical thermodynamics reflects some objective features of the system---it is large, it is near equilibrium, etc. Pragmatic factors may thus shift the focus from an internal to an external view on the duality, and dictate our preference for one of the two sides.

Finally, let us ask: what are we to make of cases where a duality is {\it broken}? These are cases where (Identical) fails: though the two theories may give the same results in some relevant approximation scheme (e.g.~perturbation theory), it is impossible to establish (Identical) beyond that approximation, i.e.~to have a duality without changing at least one of the two theories. In this case we only have an approximate duality (that is, in fact: no duality!). I will return to this case in the next section, when I discuss emergence.

\section{Gravity and Emergence}\label{emergent}

There has been a proliferation of papers on `emergent space-time' and `emergent gravity', related to dualities, in recent years.\footnote{See e.g.~Carlip (2014), Teh (2013).} The intuitive idea is this: if theory F and theory G are dual to each other, and theory F is some kind of field theory with no gravity whereas theory G is a theory of gravity; then there must be some suitable sense in which gravity emerges in theory G from theory F: the latter being regarded as more fundamental. 

There clearly {\it is} a connection between dualities and emergence; but it is {\it not} the one just quoted. In this section, building on the analysis in Dieks et al.~(2015), but using the new framework for dualities introduced in section \ref{cond}, I will spell out the kind of relationship that there {\it can} be between duality and emergence: namely, there can be emergence when there is {\it approximate duality}. From this relationship, two kinds of emergence will be seen to be possible: and I will discuss several examples, including AdS/CFT, Verlinde's scheme, and the putative emergence of five-dimensional black holes at the RHIC experiment. 

In section \ref{fall}, I will argue that emergence can only relate to {\it approximate} dualities, the duality being modified by {\it coarse graining}. In section \ref{emapprox}, I will spell out, based on the framework of section \ref{cond}, where coarse-graining can take place---i.e.~I will identify the sources of breaking of duality, and discuss several examples. I will introduce Verlinde's scheme in section \ref{Vscheme}. Then in section \ref{BH}, I will provide a new derivation of the black hole entropy formula and discuss its interpretation and significance for Verlinde's scheme. I will address the issue of emergence in Verlinde's scenario in section \ref{emgr}, where I will also get back to the emergence of black holes. Finally, I will take up the question of background-independence in the context of Verlinde's scheme (section \ref{biV}).

\subsection{The asymmetry of emergence}\label{fall}

Let me spell out how, and why, emergence cannot take place when there is an exact duality. 
%In section \ref{fall} I reviewed an obstruction to this statement: if F and G are dual descriptions of the whole world, then there does not seem to be a sense in which one theory can be more fundamental than the other. They are different formulations of the same theory describing the same physics, even if their interpretation may be rather different---this is the internal point of view reviewed in section \ref{interp}.
The following line of reasoning is fairly often found in physics papers (for references, see section 3.3.1 of Dieks et al.~(2015)):
%I give an example here: consider the following argument, which I take from Dieks et al.~(2015) but which I here write in a more abstract form:
\\
\\
(a) theory F and theory G are dual to one another; \\
(b) an exact formulation for theory F is known; but such a formulation for G is not known; therefore: \\
(c) theory F is more fundamental than theory G. \\
\\
This argument is fallacious because, starting from the premise of duality, it appeals to the pragmatic situation that an exact formulation of G is not known, to conclude that theory G is less fundamental. But this contradicts the premise that theories F and G (rather than their currently known formulations) were dual to one another. After all, an exact duality implies a one-to-one relationship between the values of all possible physical quantities, so that it seems wrong to claim that one of the two theories is more fundamental, basic, or theoretically superior. Since (b) does not imply that a fundamental formulation of G does not exist, the inference from (b) to (c) is invalid. For all we know, a fundamental formulation of theory G might be found tomorrow!

More precisely: duality is a symmetric relation: if F is dual to G then G is dual to F; whereas emergence is {\it asymmetric}: if F emerges from G, then G cannot emerge from F; and {\it non-reflexive}: G cannot emerge from itself. Thus, the presence of a duality relation---in the absence of additional relations---{\it precludes} emergence.

Consider the following objection. A metaphysical realist might hold theory F to be more fundamental than theory G because its variables stand in a truer (perhaps: the only true) relation to the world. In this way, the metaphysical realist might claim to have made room for emergence. But this objection is misguided, since  it does not introduce an appropriate relation between the system and its comparison class (cf.~beginning of section \ref{emapprox} for how such a relation is needed on the conception of emergence I will adopt). Metaphysical realism may favour the external over the internal viewpoint but it does not give us emergence: breaking the symmetry between F and G is a necessary but not sufficient condition for emergence. What is needed is an approximative relation that allows for a comparison between  the system and the appropriate class. An {\it interpretation} of a theory can {\it not} by itself lead to emergent properties or behaviour.

The next section highlights the usefulness 
%added
of careful thinking through dualities. Fallacies like the one above make for bad heuristics for physics: if theory F is more fundamental than theory G then there is no point in looking for a fundamental description of G. And so, reliance on a mistaken conclusion might result in missing interesting physics!

\subsection{Emergence from approximate dualities}\label{emapprox}

If, as argued in the previous section, duality precludes emergence, then where should one {\it look for} emergence? What truth is there in the literature's suggesting the existence of a close relationship between duality and emergence?

I will take emergence to be ``properties or behaviour of a system which are novel and robust relative to some appropriate comparison class'' (Butterfield  (2011), section 1.1.1). The relevance of duality to emergence is that it provides the relevant comparison class. But, as argued, the duality needs to be modified in order for emergent properties or behaviour to arise: so that there can be novelty and robustness (or autonomy). These two seemingly contradictory requirements---duality precludes emergence; but duality is expected to provide the relevant comparison class---can be reconciled by having an {\it approximate duality}. On the one hand, the duality is modified by an approximation; but the approximation, being a matter of degree, allows the two theories to be `close enough' (by the putative duality) that a quantitative comparison can be made. Thus, an approximate duality satisfies all the desiderata for emergence.

Let me describe in more detail how a duality will be modified. The approximations, made to describe particular physical situations or systems, will consist in coarse graining over irrelevant degrees of freedom. Coarse-graining is measured by a parameter (or family of parameters!) that can be either continuous or discrete: but there should be a notion of `small' steps, so the approximation is sufficiently accurate to describe the situation or system at hand. It was argued in Dieks et al.~(2015) (sections 3.2.2-3.2.3) that gravity can indeed emerge after such {\it coarse-graining},\footnote{Interesting accounts of emergence via coarse-graining, in the context of RG and effective field theories, include Butterfield and Bouatta (2015), Crowther (2015).} for instance in a thermodynamic limit; and that this accounts for both the novelty and the robustness of the emergent behaviour. This argument can now be generalised, based on the conditions for an exact duality given in \ref{cond}.

If emergence appears in coarse-grained dualities, we can precisely identify, from the duality conditions in \ref{cond}, where coarse-graining can take place that is suitable to emergent behaviour and properties. As I now argue, such coarse-graining can involve either one of two situations.

\subsubsection{Two ways for coarse-graining to yield emergence}

The two ways for coarse-graining to yield emergence correspond to two different modifications of the duality conditions in section \ref{cond}:\\
\\
{\bf(BrokenMap)} The duality map (Identical) breaks down at some level of fine-graining: so that the map fails to be a bijection (it is bijective only in some perturbative sense). Thus there is only a {\it perturbative duality}: one that breaks down at some order in perturbation theory or other approximation scheme; there is no duality of fine-grained theories. In this case, if theory G provides a description of the world to some accuracy, and G is perturbatively dual to F, then we can say that (aspects of) the behaviour and physical quantities described by theory G emerge, by perturbative duality, from theory F. This is how gravity (and space) emerge in Verlinde's scheme (cf.~section \ref{emgr}). This is also the situation discussed in paragraph \ref{pragma}: a specific kind of failure of (Identical).  

Notice that emergence as (BrokenMap) applies independently of the condition (Consistent). The two theories F and G are (Consistent) (with respect to {\it some} relevant body of phenomena) while {\it inequivalent}: the approximate duality map should not extend to the fine-grained level, on pain of becoming full duality and destroying emergence. The most straightforward case is the one in which the emerged side of the duality, G, does not extend to a (Consistent) fine-grained theory, while F does extend to such a theory. The broken duality then relates the fully (Consistent) F to a theory G that lacks a fine-grained definition.\footnote{This is the second view on the status of dualities as theories of quantum gravity, discussed in the Introduction (section \ref{twov}). It is the guiding idea in several of the extant quantum gravity programmes.} The world seen through the lens of G is a hologram of limited resolution: a fine-grained description of the world would require use of the more fundamental F.\\
\\
{\bf(Approx)} An {\it approximation scheme} is applied on each side of the duality (each of the theories), so that now the dynamical laws and-or quantities only describe the systems approximately. (Approx) introduces an approximation within given theories ({\it families} of theories) related pairwise by exact duality, at each level of coarse-graining. Duality and coarse-graining thus work on different directions.  The condition (Consistent) now only holds approximately; and its set of physical requirements may be replaced by a new set, appropriate to the approximated situation. Emergence happens because there are structures and physical quantities (temprature, or the gravitational force), or behaviours (increase in entropy, or the universality of gravity) that: (i) did not figure in the fine-grained theory (the theory out of which theory G emerges by coarse-graining); but that (ii) do apply after the approximation (usually involving a limit) is applied. This is the situation discussed in section \ref{interp}, just before \ref{pragma}: a shift in the relevant body of phenomena stipulated in (Consistent). 

At first sight, the second kind of emergence might seem uninteresting: whatever emergent properties or behaviour are on one side, are mapped to other emergent properties or behaviour on the other; but the emergent behaviour itself has nothing to do with duality. (Approx) becomes much more interesting, and a helpful tool for {\it producing} emergence in cases of duality, when we consider dualities with external parameters or boundary conditions\footnote{In what follows, I will call both generically `parameters'.} that can be freely varied. Varying these parameters, a series of appoximations can be produced that can exhibit the emergence of new phenomena or behaviour, while retaining duality {\it at each level}: thus, we are considering a {\it series} of dualities! Here, all emergence takes place on one side of the duality: and whatever emergence there is in G, is mirrored in F by the duality: even if it takes a completely different form! (Approx) thus preserves (Identical) at each level, though there is a different bijection for each value of the external parameters. 

Because the duality scheme in section \ref{cond} has three conditions, there is a {\it third} way duality may fail: a lack of (Num). A subset of the quantities between the two theories may agree; but the two sets of quantities are unequal in number. A case in point are topological string theories, which calculate a subset of the quantities of the full, `physical' string theories (namely, quantities that can be computed using topological or geometrical arguments).\footnote{I do not claim that topological string theories {\it only} calculate a subset of the physical quantities, or that they agree with that subset in {\it all} cases: only that a subset of the quantities {\it does} agree.} It is hard to see, however, how concentrating on a subset of the physical quantities could lead to emergence if the agreement between the quantities in that subset is otherwise exact. Taking a subset out out of all the quantities, there is only a notion of belonging to that set or not; but no notion of successive approximation such that robustness and novelty of behaviour arises: there is no coarse graining. Thus, these are not cases of emergence but of idealization by restricting to a subsector. 

\subsubsection{General comments and comparison}

Following the discussion in section \ref{fall}, where it was argued that emergence can only appear if duality is broken, I have now identified two modes of emergence as failures of one of the conditions in my duality scheme in section \ref{cond}. And because the third duality condition was argued {\it not} to produce emergence, this shows that these two possibilities {\it exhaust} the available options; on the notion of emergence quoted at the beginning of this section: from Butterfield (2011). Of course, the above two modes of emergence can be combined.

In Dieks et al.~(2015), emergence in AdS/CFT and Verlinde's scenario was discussed. The present framework construes the conceptual difference, between these two types of emergence, as a distinction in the specific {\it duality condition} that is modified. This allows for the framework to be applied to other dualities.

Both cases of emergence are cases of coarse-graining. Emergent behaviour, or emergent quantities, are usually found after a limit is taken; however, as remarked in Butterfield (2011), also {\it before} the limit is taken, emergent behaviour can already arise. Indeed, one should be cautious: there is not always a {\it continuous} parameter that is being varied in the coarse-graining approximation (taken to zero, or to infinity). This parameter may well be discrete (in AdS/CFT, one of the relevant parameters, the rank of the gauge group $N$, is a natural number taken to infinity: cf.~the next subsection), and coarse-graining may take place in discrete steps (as as in block spin transformations, where degrees of freedom are integrated out in successive block spin transformations).

As we saw in section \ref{cond} in the discussion of Newtonian gravity, theories with {\it limited applicability} can feature in dualities.\footnote{This is the case in (Approx), where duality obtains at each level of coarse-graining. The coarse-grained theories can have duals; they nevertheless have limited domains of applicability.} The condition on these theories is that they be (Consistent) with respect to the relevant body of physical phenomena and within a prescribed mathematical domain of applicability. This guarantees that in (BrokenMap), though there is no fine-grained duality, there can still be an exact duality between coarse-grained theories: between F' (obrained from F by coarse-graining) and G; and the properties or behaviours in G emerge from F not F'.  

In (Approx) emergence, all emergence takes place within G. If one claims that gravity emerges in G after coarse-graining, then  the meaning of `gravity' needs to be explicated: since there is a claim of gravity's {\it absence} in G at the fine-grained level. Thus, a precise conception of the expression `a theory of gravity' is needed. For instance, should string theory or M-theory, at the fine-grained level, turn out to be theories of gravity (by whatever microscopic concept of `gravity' is adopted), then one should only say that {\it Einstein gravity} (general relativity) emerges as an approximated theory of space-time.

As I will argue in section \ref{emgr}: Verlinde's scheme is indeed a case of (BrokenMap); and therefore it does not face this challenge. That is: there is no fine-grained theory G, and gravity appears as a holographic reformulation of the degrees of freedom of theory F in the thermodynamic limit.

\subsubsection{Applications of the two modes of emergence}

(Approx) explains how general relativity can emergene in AdS/CFT: the string theory description is assumed to be dual to the quantum field theory; but for a certain range of parameters (string coupling and number of D-branes $N$) general relativity (and its properties: Einstein gravity) emerges as an effective theory. Of course, this discussion is mirrored by an appropriate phenomenon of emergence on the quantum field theory side, too. The important point is that the original exact duality has been replaced by a series of dualities, one of them now relating general relativity to a strongly-coupled quantum field theory.

As a second example, let us consider the RHIC experiments carried out in Brookhaven, NY: to which gauge/gravity duality was succesfully applied. Here, a five-dimensional black hole is employed to perform a calculation that, via approximate duality, provides a result in a four-dimensional gauge theory (the shear-viscosity-to-entropy-density ratio of a plasma) that can (at present) not be obtained in the theory of QCD  describing the plasma (or in any other theory, for that matter). Since there is no exact duality between gravity and QCD, this might seem a putative case of (BrokenId). 
Does this mean that a black hole `emerges' from the RHIC experiment? No: there is no emergence of gravity here because the black hole is on the wrong end of the relation of emergence. The emergence relation is asymmetric (section \ref{fall}), hence it obviously matters which is the approximated theory.Theory G (the gravity theory with putative (BrokenMap) emergence) does not describe actual features of the world (which are described by theory F instead, i.e.~QCD). G is just a calculational tool: its gravitational entities are fictitious. We may speak of `effective gravity' or an `effective black hole' but not of emergence: since the black hole has no physical reality: there is no corresponding `system' to which emergent properties or behaviour can be attributed (see the beginning of this section).

Having analysed how emergence can arise when dualities are approximate, and applied this in two examples, I will devote the rest of this section to my third and main example: emergence in Verlinde's scheme. I will first review, in subsection \ref{Vscheme}, Verlinde's derivation of Newton's law based on holography and thermodynamics, emphasising several conceptual aspects that were not discussed in Dieks et al.~(2015). In subsection \ref{BH} I will give a novel derivation of the Bekenstein-Hawking entropy formula based on Verlinde's scheme. This will allow us to go back, in section \ref{emgr}, to the issue of emergence of black holes, which now {\it are} related to emergence; along with an analysis of emergence of gravity generally in Verlinde's scheme.

\subsection{Verlinde's scheme}\label{Vscheme}

In this section I introduce Verlinde's scheme for getting Newton's law from thermodynamics. The derivation given here is my rendering of the original Verlinde (2011), also analysed in Dieks et al.~(2015). With respect to the latter, I will here: (i) Make a clear distinction between the reservoir and the system, their interactions, and their interpretations in the bulk. (ii) I will discuss in more detail the relation between motion in the bulk and RG (renormalization group: see section \ref{notation}) flow. Finally, I will: (iii) Make one of Verlinde's assumptions explicit, namely Shannon's conection between entropy and loss of information.

\subsubsection{Setup: system and reservoir}

Consider a `system' of $n$ bits coupled to a reservoir at temperature $T$ and with a total energy $E=\half NkT$. $N\gg n$ is the number of bits in the reservoir; and knowing the total energy of the reservoir and the number of bits one knows the temperature. Therefore the reservoir is characterized by two numbers, $E$ and $N$. In this composite system an entropic process takes place by which the bits tend to an equilibrium. It is this entropic process that will give rise to gravity.

The quantities that characterize this system, together with its reservoir, are therefore $E,N,n$: the first two for the reservoir, $n$ for the system itself. To set up a holographic relationship between the composite system and the bulk, we have to imagine the composite system as being embedded on the `boundary' of some space, e.g.~a spherical screen of area $A$ enclosing a total mass $M$, in a flat space outside.\footnote{$n$ will correspond to the mass just outside the screen: see footnote \ref{nbits}.} The bulk and the reservoir are mapped to each other through the following relations:
\bea\label{ETN}
N&=&{A\,c^3\over G\hbar}\nn
{E\over c^2}&=&M~.
\eea
I stress that at this point this is the {\it definition} of the holographic map between: on the one hand, space and mass; and on the other, the reservoir (its number of bits, and energy). The numerical constants are chosen for dimensional reasons; being definitions, their numeric value is irrelevant (see subsections \ref{ecg} and \ref{discbh}).

\subsubsection{Emergence of space as a coarse-graining variable}\label{ecg}

Further, let us imagine that the bits $N$ and $n$ are of the same nature; because $N$ is large, a description in terms of equipartition and temperature is applicable. For instance, we can think of an Ising model starting at some fiducial temperature and ending up at temperature $T$ after a number of block-spin transformations---of averaging over spins and rescaling (see e.g.~Chandler (1987)). 

In the case at hand, this RG flow will be represented by a trajectory in the space labelled by a parameter $r$. $r$ labels the level of coarse-graining in the boundary theory and can now be used to distribute the composite system with its $N+n$ degrees of freedom radially, as a series of concentric spheres related by coarse-graining operations. 

Consider a particle approaching the screen and interacting with it: this is dual to the system we describe in interaction with the reservoir. From the outside, the particles that went inside the screen cannot be seen; but some of their effects can be felt: just like the reservoir, whose microscopic degrees of freedom cannot be probed: but it can exchange pressure and heat with the system (this is the meaning of `interaction', used above). Recall Bekenstein's thought experiment. When the particle is well outside the screen, its position can be tracked: but when it reaches one Compton wave-length from the screen, it can no longer be distinguished from it. Because of the duality we have stipulated, between the screen (with whatever went in) and the reservoir: it follows that the screen is not part of the space-time accessible to particles (the system): the screen is a boundary, the part of space-time that is not yet observable from the outside---it has not yet emerged. So when a particle is less than one Compton wave-length from the screen, its position can no longer be tracked; the particle must now be dual to some bits that belong to the reservoir (by energy conservation). The fine-grained details of the particle are irrelevant to coarse-grained physics---the relevant effects are captured by increases in entropy. Throwing particles in thus amounts to removing particles from the system and adding them to the reservoir, thereby removing some information from the system, which results in a change of thermodynamical state. Removing particles from the system, adding them to the reservoir, is analogous to the operation of `integrating out' in the Ising model: in Verlinde's scheme\footnote{Based on AdS/CFT intuition, where motion in the additional dimension has indeed plausibly been shown to be dual to RG flow in the quantum field theory (see section \ref{notation}, fourth bullet).}, it is dual to an RG operation in the fine-grained theory.

I just argued that $\Delta r$, the distance to the screen, is dual to a variable that measures an (infinitesimal amount of) RG flow. But one should recall that, in general, RG flow involves two operations: 1) lowering the cutoff together with integrating out degrees of freedom above that cutoff, 2) rescaling of variables so that the new cutoff plays the same role as the old one.\footnote{For AdS/CFT, this was discussed in some detail in Dieks et al.~(2015).} In the context of Verlinde's scheme, I have discussed integrating out, i.e.~step 1), in the previous paragraph, where a particle was thrown in. I now turn to rescalings, operation 2).

When a particle is added to the reservoir, $N$ increases by one. The area \eq{ETN}, however, can be held fixed if a simultaneous rescaling of $\hbar$ is carried out. The identification between bulk and boundary quantities \eq{ETN} indeed involves the constant $\hbar$, which is introduced for dimensional reasons. This proportionality constant is part of the {\it definition} of the holographic map---of how the boundary theory is embedded in the bulk. Hence $\hbar$ is not an invariant physical quantity but a choice of length scale. Indeed, it drops from the final formula for the force. 
%added

Thus, there is a notion of rescaling of variables under which \eq{ETN} takes the same form and the force is invariant---so Verlinde's identification of small displacements with RG trajectories makes sense. The next step is to calculate the entropy. The key point is to quantify the amount of information lost.

Consider Figure \ref{Verlinde-figure1} below. When the particle advances by a distance $\Delta r={\hbar\over mc}$ (the particle's Compton wave-length) from the red to the green screen, it disappears behind the red screen and information is lost from the bulk. The amount of information lost is estimated by Bekenstein to be roughly one bit of information, in conveniently chosen units:
\bea
\D I_{\tn{lost}}=-2\pi k_{\tn{B}}~.
\eea
Using Shannon's connection between loss of information and increase in entropy, the entropy increases by an amount $\D S=-\D I_{\tn{lost}}$:
\bea\label{entropy}
\D S=2\pi k_{\tn{B}}=2\pi k_{\tn{B}}\,{mc\over\hbar}\,\D r~.
\eea
\begin{figure}[here]
\begin{center}
\includegraphics[height=5cm]{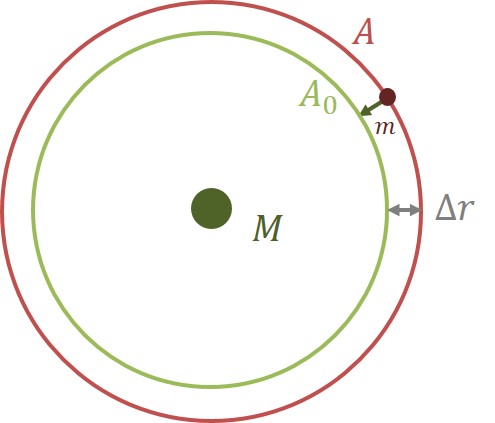}
\caption{\small{A test mass falls into the screen.}}
\label{Verlinde-figure1}
\end{center}
\end{figure}

As we have just seen, from the bulk point of view, $r$ is the position of the particle falling in; from the point of view of the boundary, it is a book-keeping device that records the level of coarse-graining and hence the increase in entropy as more information is lost. The force associated with this increase in entropy is given by the second law of thermodynamics: $F=T\,{\D S\over\D r}=2\pi k\,{mc\over\hbar}\,T$. Solving for $T$ from \eq{ETN} and writing the area of the sphere $A=4\pi R^2$, we get Newton's law:
\bea\label{Newton}
F=G\,{Mm\over R^2}~.
\eea
I will say more about the physics of this equation in the next section. 
%But first let us turn to the philosophy of emergence in  \ref{emgr}.

%If we rescale the sphere to a sphere of smaller radius $A\rightarrow A_0=\l\,A$, keeping $N$ and $T$ fixed, then \eq{ETN} is fixed if we also rescale $\hbar\rightarrow\ti\hbar=\l\,\hbar<\hbar$. 

\subsection{The Bekenstein-Hawking entropy formula}\label{BH}

Verlinde does not discuss black holes, though his argument for particles approaching screens, explored in section \ref{Vscheme}, clearly mimicks Bekenstein's original thought experiment. Bekenstein {\it was} concerned with black holes and their entropy; and his result was correct up to a dimensionless constant that he estimated, but came out too small by a factor of ${2\pi\over\ln2}\approx9.065$. Estimating that number entailed explicit use of general relativity.\footnote{In appendices A and B of Bekenstein (1973).}

On the other hand, Verlinde's derivation so far did not invoke black holes or general relativity, other than as general motivation.\footnote{Verlinde's paper (section 5) contains a generalization of his scheme to get general relativity from screens.} It is therefore natural to ask whether: (i) Verlinde's scheme can calculate black hole entropy, (ii) whether it fares better than Bekenstein's entropy?

In this section I will show that the answer to both questions is positive: assuming the coarse-graining variable to take the form of Newton's potential, one finds exactly the Bekenstein-Hawking entropy, including the correct numerical factor, and without the need to invoke general relativity. From the interpretation of Newton's potential as a coarse-graining variable, one also finds an interesting interpretation of the black hole horizon: in terms of RG flow. I will discuss the implications in the next section.

\subsubsection{Derivation of black hole entropy}

In this subsection I will give a novel derivation of the black hole entropy formula based on Verlinde's arguments. In the discussion so far in section 3, we have been concerned with the thermodynamical properties of the reservoir. Let us now describe the system itself in thermodynamical terms. In analogy with \eq{ETN}, we have\footnote{\label{nbits}The assumption here is that a particle of mass $m$ corresponds to $n$ bits (the `system'), each with energy $\half k_{\tn{B}}T$. As long as the particle is outside the screen, the bits of the system are well distinguished from those in the reservoir. This justifies the earlier statement that the system is dual to the bits {\it just outside} the screen (within approximately one Compton wave-length from it), whereas the reservoir contains the bits of the particles that already fell into the screen.}:
\bea
m={1\over2c^2}\,n\,k_{\tn{B}}T
\eea
as well as:
\bea
{\bf a}=-{\bf\nabla}\F~,
\eea
where $\F=-{GM\over r}$ is Newton's potential, the variable that measures the level of coarse-graining. For small displacements $\D r<0$ (inward fall) we also have $\D\F<0$, hence the acceleration points inward and
\bea
{\D S\over n}=k_{\tn{B}}\,{\D\F\over2c^2}>0\label{DS}
\eea
increases as the particle falls in. Thus the direction of increasing coarse-graining is indeed the direction of decreasing Newtonian potential, $\F<0$. 

Let us consider what happens when we reach the maximum potential, the potential at the horizon of a black hole: $r=R_{\tn{S}}={2GM\over c^2}$. In that case:
\bea\label{maxpot}
\F(R_{\tn{S}})=-{c^2\over2}~~\Rightarrow~-k_{\tn{B}}\,{\F(R_{\tn{S}})\over2c^2}={1\over4}\,k_{\tn{B}}~.
\eea
In particular, consider a particle falling in from infinity $\F=0$ towards the horizon. Its change in entropy is:
\bea\label{deltaS}
\D S={1\over4}\,k_{\tn{B}}\,n~.
\eea
Let us now assume that the entire entropy of the black hole was built up by the repetition of such a process: particles coming in from infinity, each of them adding an amount of ${1\over4}\,k_{\tn{B}}$ per bit. Using \eq{ETN} for these $N$ particles, we get exactly the Bekenstein-Hawking entropy formula:
\bea\label{Sentropy}
S={1\over4}\,k_{\tn{B}}\,N={1\over4}{k_{\tn{B}}A\,c^3\over G\hbar}~.
\eea
The novelty here is that we get the black hole entropy with the correct factor of 1/4 even though the normalization \eq{ETN} did not include such a factor. Furthermore, \eq{ETN} was a {\it definition} of the number of states, whereas \eq{Sentropy} calculates something physical. 

\subsubsection{Discussion of black hole entropy}\label{discbh}

The normalization in \eq{entropy} was chosen to reproduce Newton's law and, as the derivation \eq{DS}-\eq{Sentropy} shows, Newton's law contains the information about the normalisation of the black hole entropy formula. The factor of 1/4 in Bekenstein's formula follows, in Verlinde's scheme, from the same principles as Newton's law, without using Bekenstein's black hole calculations (Bekenstein (1973): appendices A and B). This consistency check adds to the robustness of Verlinde's derivation of gravity. 

The derivation sheds light on a further aspect of Verlinde's scheme: the chosen constants $\hbar$ and $c$. These were introduced in \eq{ETN} for dimensional reasons, and, as I discussed in paragraph \ref{ecg}, they defined the holographic map; their value did not affect the final result: Newton's force, obviously, does not depend on either of these constants! However, their value enters the final physical result \eq{Sentropy} for black hole entropy. So what just happened? 

Let us see how the constants are interpreted in the thought experiment in the previous subsection. After the initial Eq.~\eq{ETN}, the {\it second} appearance of $\hbar$ is in the Compton wave-length of the mass $m$, which precisely cancels with the $\hbar$ in \eq{ETN} in the formula for the entropy \eq{deltaS}. This is a matter of setting the scale: the holographic map \eq{ETN} introduces a notion of length, which then determines the wave-length. These two notions are correlated and can be varied simultaneously. But then this also sets the scale for a further holographic formula for $n$: integrating \eq{deltaS} to deliver the entropy \eq{Sentropy}. As pointed out, the assumption was that the bits $n$ in the system were of the {\it same kind} as those in the reservoir, $N$. Indeed: once the first holographic map \eq{ETN} is set, $\hbar$ (together with other constants) determines the bits per area: this is a matter of how one embeds the reservoir at the boundary. A different holographic map would give a different Compton wave-length and holographic formula for $n$. A similar argument applies to $c$. In other words, rescalings in \eq{ETN} will induce a redefinition of the notion of length, hence in how the area is measured in \eq{Sentropy}: and the final result is invariant under such rescalings.

From the bulk point of view, $\hbar$ (up to constants) is interpreted as the typical wave-length corresponding to a particle of unit mass. From the dual point of view, this wave-length is the infinitesimal length of the shell that is integrated out: the {\it distance between two successive cutoffs in RG flow}. As in RG, the choice is immaterial: successive RG steps require rescaling of this length.

The interpretation of the speed of light $c$ is enlightening. It appears in the formula for the Schwarzschild radius, which is fixed by the minimum of the gravitational potential $\Phi(R_{\tn{S}})$: the maximum value of the potential difference. As we see from \eq{maxpot}, $c^2/2$ is this maximum value. The interpretation of $c^2/2$ is then as the maximum level of coarse-graining in the boundary theory, i.e.~the {\it coarse-graining at the fixed point of the RG}. How this coarse graining is measured is of course fixed by the variable chosen. But once $\F$ is taken to be Newton's potential, the maximum value of the potential difference is physical. 

Chirco et al.~(2010) have noted that the numerical factor of 1/4, appearing in the formula for the entropy, is related to the equivalence principle (assuming a relativistic description). Though there are links between Verlinde's general scheme for getting general relativity holographically and e.g.~Jacobson's thermodynamic description, the particular derivation given here does not assume the equivalence principle nor relativity. In particular, $c$ does {\it not} have the interpretation of the speed of light in the boundary theory (see the previous paragraph): which, indeed, need not be a relativistic theory, nor a theory of space: see section \ref{biV}. In Verlinde's scheme, the concept of inertia is emergent: a particle is at rest because of the absence of entropy gradients. The equivalence principle arises, after holographic reformulation, as the ability to choose a screen (a coarse-graining scheme) in which, for a system small enough, there is no entropy gradient. 

\subsection{Black holes, AdS/CFT, and emergence}\label{emgr}

We can now relate the results in sections \ref{Vscheme} (Newton's force: Eq.~\eq{Newton}) and \ref{BH} (black hole entropy: Eq.~\eq{Sentropy}) back to the analysis of emergence in approximate dualities performed in section \ref{emapprox}. For comparison, I first discuss AdS/CFT.

Emergence of gravity in AdS/CFT can be construed as a case of (Approx). The nature of the approximation is the large $N$ limit, in which (near a fixed point) RG flow is suitable for describing certain situations (e.g.~low energies, see section \ref{notation}); and so it is akin to taking a thermodynamic limit. Now in such a scheme, the property (Identical) at each level of coarse-graining is inherited from the fine-grained duality: but there is a modification of (Consistent) in that, at each level of coarse-graining, the relevant class of phenomena changes. This explains why, and how, emergence in AdS/CFT is a case of (Approx). 

If so, and regarding these theories as (Whole) so that one is justified (or takes oneself to be justified!) in adopting the internal point of view (ii) of section \ref{interp}, there is no reason why one side should be more fundamental than the other. Rather, it is the {\it thermodynamic limit}, as a modification of (Consistent), that introduces the possibility for gravity to emerge on one side. 
%In such a case, one has to be more precise about what one means by a `theory of gravity', as remarked in option (ii) at the beginning of section 3. 

On the other hand, Verlinde's scheme, as construed here, does not preserve (Identical) at all levels---the duality map only exists after coarse-graining---hence it is a case of (BrokenMap). There is no (Consistent) quantum theory in the bulk; there is only one fine-grained theory F, which is embedded on the sphere via the holographic map, but its details are irrelevant; only its universality class matters. In the thermodynamic limit, this theory flows towards a coarse-grained theory characterised by $E,N,T,n$, dual to the gravity theory in the bulk. It is here that gravity can emerge from F as in (BrokenMap). In conclusion, the crucial ingredient in this kind of emergence is not holography, but the existence of a thermodynamic limit of theory F, which I called theory F' (section \ref{emapprox}).

Assuming a specific coarse-graining function corresponding to Newton's potential, black hole entropy can be calculated in Verlinde's scheme to give the correct value. The Schwarzschild radius corresponds to the maximum level of coarse graining. Some of the properties of the black hole (its entropy, the force on a test particle) do indeed correspond to thermodynamical quantities obtained via coarse-graining: they are {\it emergent}. 

How does this differ from the RHIC case, in which I explained that the black hole did {\it not} emerge and was fictitious? The difference is clear: the black hole here is an ordinary, four-dimensional black hole; the black hole at RHIC was five-dimensional and on the wrong side of the relation of emergence. None of the two were exact duals of a (Consistent) theory; and precisely for that reason---there is no duality---there is only an external point of view: only the four-dimensional black hole, in so far as it is empirically adequate, is considered physical. The external point of view and the asymmetry of emergence work together to select one of the two black holes as emergent and the other one as effective.

\subsection{Background-independence in Verlinde's scheme}\label{biV}

In section \ref{Vscheme} I introduced Verlinde's scheme as a {\it duality} between a statistical mechanical system coupled to a reservoir; and a particle near a screen containing a certain amount of mass and interacting with it. The position $r$ of the particle was argued to be dual to the coarse-graining variable (that effectively splits system from reservoir); and $\hbar$ and $c$ had interpretations as the distance between two cutoffs, and the measure of maximal coarse-graining, respectively. Verlinde's scheme is a case of (BrokenMap) and the sense in which it is appropriate to speak of a duality was explicated in section \ref{emapprox} and \ref{emgr}: duality obtains between theory G (general relativity, with its specified domains of applicability) and theory F' (the coarse-grained version of the theory F, embedded on the screen), though it does {\it not} extend to a fine-grained duality between G and F. In this section I discuss the implications of the duality between G and F'. Since Verlinde claims that G is general relativity, background-independence must apply to it: and the question is whether this property is indeed compatible with the envisaged duality. Since I have only discussed in detail how Verlinde's scheme lets Newton's law emerge, but not how general relativity emerges, I will limit myself to making one suggestion: that Verlinde's scheme best fits the extended sense of background-independence.

In section \ref{extended} I contrasted  minimalist vs.~extended senses of background-independence; the former is a conservative notion modelled after general relativity; the latter takes what is most characteristic of Einstein's theory and promotes it to the status of a heuristig guiding principle for the construction of theories of quantum gravity, even beyond the form background-independence takes in general relativity. 

Verlinde's scheme can be interpreted conservatively, such that the minimalist sense obtains: one starts with a two-dimensional {\it surface} (i.e.~fixed geometry) with bits on it, etc.\footnote{Verlinde's scheme was introduced in Dieks et al.~(2015), in section 4.1, as follows: ``Imagine a closed two-dimensional space, e.g. the surface of a sphere, on which a quantum
theory is defined.''} Geometry is already present on the screen in a way reminiscent of the standard presentation of AdS/CFT, in which the asymptotic boundary geometry is fixed: as discussed in section \ref{adscftc} (but see the comments on the extended sense at the end of \ref{extended}). 

There is a more adventurous interpretation of Verlinde's scheme that does not start with the surface of a sphere, but with a theory in which there is no space: for example, an Ising model on a lattice---where the lattice should really be thought of as an abstract structure with no space. The embedding equation \eq{ETN} is then the {\it definition} of the holographic map, as noted earlier. On the left-hand side there are no space-time variables: only bits and their energies. On the right-hand side are area and mass: {\it bulk} quantities. Thus, two points should be made: (i) Eq.~\eq{ETN} is really a way to embed: or, better, reformulate: a theory in which there is no space; as a theory where some quantities are interpreted as areas and lengths. As discussed in section \ref{discbh}: the map defines the units in which distances and time-scales are measured. (ii) The screen can take {\it any} shape, as is shown in section 4.1 of Verlinde (2011). Both points hint at the {\it extended} sense of background-independence as underlying my construal of Verlinde's scheme.\footnote{One of the less clear aspects of Verlinde's derivation of general relativity (Verlinde (2011), section 5) is whether only space-times that have a timelike (or null, in Jacobson's case) Killing vector can emerge; and how time variables that differ by a diffeomorphism are to be treated on the dual side.} Of course, there is still work to be done here; but this suggests that some form of extended background-independence obtains.

The above prompts a second way (in addition to the one mentioned in the last paragraph of section \ref{extended})\footnote{Allowing boundary conditions to be selected dynamically, or even summing over boundary geometries.} in which AdS/CFT, too, might be made background-independent in the {\it extended} sense. The spatial structures in the gauge theories under consideration (such as four-dimensional super-Yang-Mills theory) might themselves arise from theories in which there is no spatial structure at all; for instance, matrix models. In fact, much of the AdS/CFT integrability programme seems to go this direction. The integrability of the planar sector (roughly: the appropriate limit of small coupling and large $N$) of the non-abelian CFT's involved, allows to calculate the relevant amplitudes from models, such as spin chains, in which there is no space-time; i.e., the spin-spin interactions are regarded as taking place on an abstract chain. Integrability features associated with discrete models have been found in several of the CFT's relevant in AdS/CFT, in various dimensions. Whether these features extend to the full CFT's  is yet to be seen, but this does suggest new avenues for the investigation of two important aspects of  gauge/gravity dualities: (i) background-independence in the extended sense; (ii) emergence of space.

\section{Conclusion}

In this paper I have analysed two main, related questions: (a) the general conditions for duality; and whether these obtain in gauge/gravity dualities; (b) the interplay between duality and emergence: emergence as {\it approximate duality}. 

On question (a): I found that dualities entail two conditions: (i) complete and consistent theories; (ii) identical physical quantities. In AdS/CFT the validity of (ii) is still open (despite a large amount of evidence that, at least in the most symmetric cases, it is satisfied). I discussed the consistency condition (i) for gauge/gravity dualities: background-independence. (ia) In the {\it minimalist} sense, background-independence is a property of the gravity theory, compatible with the need to pick (otherwise arbitrary) boundary or initial conditions to obtain a solution; (ib) the {\it extended} sense also demands background-independence of the boundary theory, i.e.~there are no boundary conditions (or these are determined dynamically). I argued that, on the account of background-independence in general relativity, only (ia) is required: because, generically, general relativity requires the supply of boundary or initial conditions that are not provided by the theory. The extended sense (ib) cannot be taken as an {\it a priori} desideratum: for standard general relativity is generically {\it not} background-independent in this sense: an argument of such an {\it a priori} status would thus weaken its own foundations. Careless use of extended background-independence might lead to throwing the baby out with the bath-water. Nevertheless, extended background-independence seems a very useful heuristic guiding principle for the construction of theories of {\it quantum} gravity that satisfy other {\it pragmatic} requirements; in particular, it is useful in finding generalisations of gauge/gravity duality: and some options were suggested.

\begin{figure}[here]
\begin{center}
\includegraphics[height=4cm]{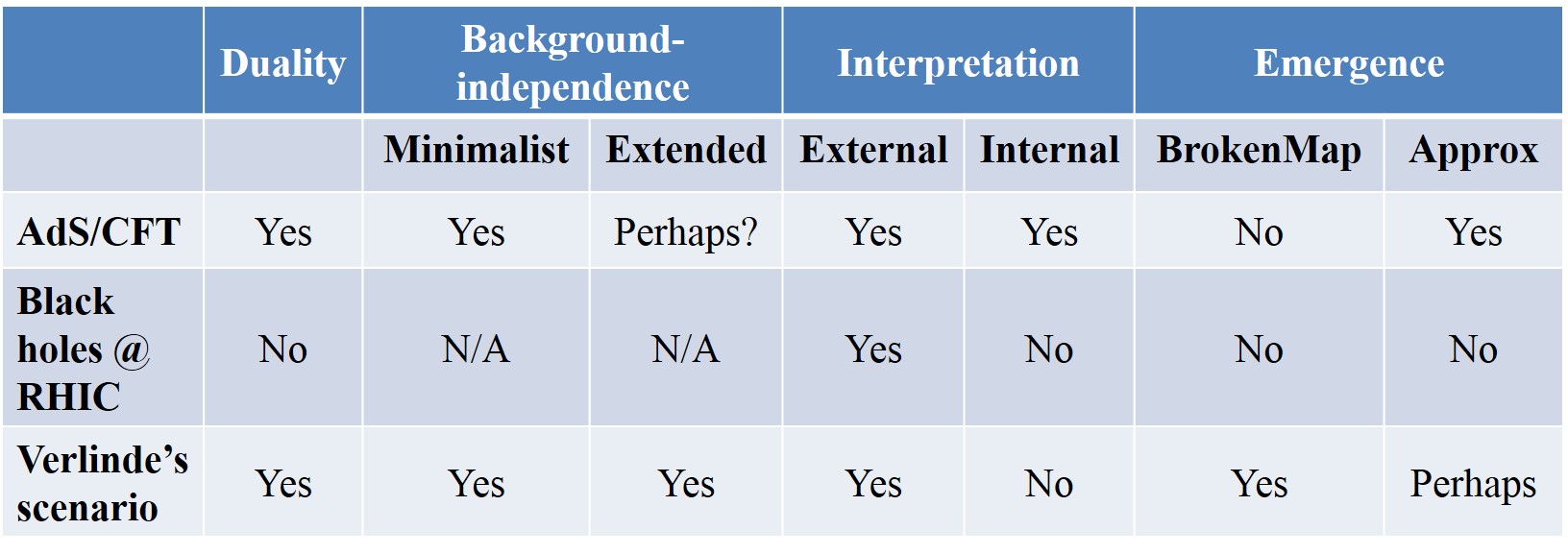}
\caption{\small{Applications of duality, background-independence, interpretation, and emergence.}}
\label{Table2}
\end{center}
\end{figure}

I applied these concepts in three main examples: AdS/CFT duality, black holes at RHIC, and Verlinde's scenario. The first is a case of exact duality; in the second there is no duality; the third is only a duality at the coarse-grained level. The findings are summarised in the table in Figure \ref{Table2}. Gauge/gravity dualities, in their current formulations, were found to be background-independent in the minimalist sense. Certainly, more work needs to be done in order to further refine and generalise the notion of background-independence; but, for the reasons explained, such refinement will not change this important conclusion if the notion keeps general relativity as the paragon of a background-independent theory.

There is an important subtlety in applying the extended notion of background-independence in gauge/gravity dualities. In the literature, it is often said that AdS/CFT fails to be background-independent because of the need to impose boundary conditions. Despite that the claim, as an expression of background-independence {\it tout court}, is wrong: the threat to the {\it extended} sense can be repaired: and I indicated two directions; whereas a larger threat has emerged in course of my analysis: (i) Indeed, one can envisage straightforward modifications that will not select a Dirichlet boundary condition (a fixed metric at infinity) but will allow a more general one, hence solving the problem of having to fix boundary conditions by hand. General boundary conditions have been studied extensively in the literature (in cases where diffeomorphism invariance is preserved: see the next point). But: (ii) the real threat to the extended notion seems to come from the presence of the {\it holographic Weyl anomaly} in {\it even boundary dimensions}. This anomaly breaks those bulk diffeomorphisms that give rise to non-trivial conformal transformations on the boundary. In particular, the holographic stress-energy tensor is not a tensor but picks up an anomalous term. This anomaly precisely matches the {\it conformal anomaly} of the CFT for curved backgrounds. The anomaly is harmless---it does not imply any kind of inconsistency of the theory (as {\it is} the case of local anomalies in quantum field theories), but, like global anomalies in quantum field theory, merely expresses the fact that some accidental symmetries of the action are not (need not) be respected upon quantising the theory. Nevertheless, the anomaly seems to be a problem if the goal is to realise {\it extended} background-independence by integrating over the boundary metrics; though the extended notion might still be achievable in the two other ways discussed: (i) changing the boundary conditions from Dirichlet to Neumann or mixed: so that a boundary condition for the metric need not be chosen; (ii) having the boundary theory be non-spatio-temporal, with an embedding into space-time that is background-independent. Clearly, this deserves more work.
%In any case, the minimalist sense seems the best one can do, in a straightforward manner, for a theory with negative cosmological constant and with Einstein gravity as its classical limit.

The conformal anomaly, and its gravity dual, the diffeomorphism anomaly, is an explicit counter-example to the claim, often made in the literature about background-independence: that covariance is cheap, since `any' theory can be made covariant, if just the right variables are chosen. As demonstrated by the anomaly, this claim is {\it false} for quantum theories as well as for classical general relativity with boundaries. In the latter case, the theory is covariant under one class of diffeomorphisms, but not under another. The upshot of this is that discussions of covariance and background-independence must take care to specify the {\it relevant} class of diffeomorphisms.

In odd boundary dimensions,  there is no such restriction, and extended background-independence seems feasible: in fact, with several options to achieve it. Extended background-independence thus points to further directions for generalising gauge/gravity dualities.

\begin{figure}[here]
\begin{center}
\includegraphics[height=3cm]{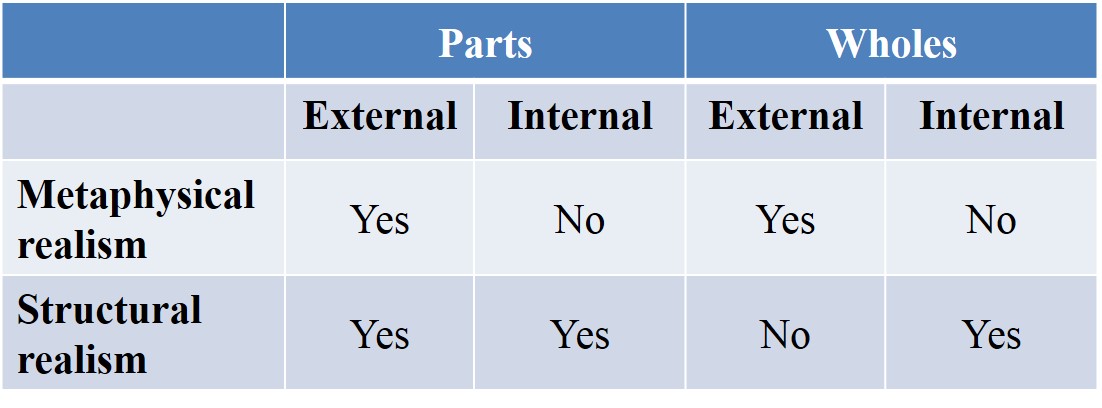}
\caption{\small{Table indicating the interpretations available.}}
\label{Table1}
\end{center}
\end{figure}

Interpreting dualities, I further spelled out the distinction between the external and internal viewpoints, introduced in Dieks et al.~(2015). The main result is that the external interpretation is argued {\it not} to be available for (Whole), other than on strong metaphysical realist assumptions. And, despite the appearance that the internal perspective might only apply to (Whole): it applies to (Parts) in cases where it can be argued that the idealisation does not disregard key aspects of the ontology of the system, such as `location'. Thus the external viewpoint is widely available on a metaphysical realist position; whereas the internal view is compatible with weaker realist stances, such as structural realism, and with anti-realist positions. This is summarised in the table in Figure \ref{Table1}.

I applied the interpretational concepts to gauge/gravity dualities: only AdS/CFT was found to possess an internal viewpoint, because from the three gauge/gravity relations studied, it is the only in which duality obtains at the fine-grained level. Black holes are found {\it not} to have emerged at the RHIC experiments in any suitable sense of the word `emergence'; instead, we should speak of `effective black holes' here. 

On question (b), of the relation between dualities and emergence: I argued that, due to the symmetry of duality and the {\it asymmetry} of emergence, the relation is one of {\it incompatibility}. The way to overcome this deadlock is to consider {\it approximate dualities}, i.e.~dualities modified by coarse-graining; where the coarse-graining could occur in two different ways, depending on which condition, (Consistent) or (Identical), is broken. This leads to two distinct forms of emergence: (BrokenMap) and (Approx). 

An important, related, finding is that the interpretation of a duality has no bearing on emergence. In particular, the external viewpoint, by itself, {\it cannot} generate emergence in any helpful sense of the word. At best, there are `effective quantities' or `effective properties': which nevertheless only describe fictitious entities; such as the black hole at RHIC. An option that my analysis also {\it excludes} is emergence by restriction of the number of states and quantities. In the presence of dualities, coarse-graining seems to be what it takes to have emergence.

In AdS/CFT, if the duality is exact, there can be emergence as (Approx).  Emergence then takes place on both sides of the duality; and duality and emergence work along different directions. Only in Verlinde's scheme did we find (BrokenMap) emergence, i.e.~(approximate) duality and emergence working along the same direction.

The two basic options: duality---without emergence; vs.: approximate duality---with emergence, reflect two distinct approaches to the problem of quantising gravity. The traditional string programme, and its current revision in gauge/gravity duality, sides with the first option; other programmes, such as analogue models of gravity and Jacobson's work, follow the second. But the present analysis reveals a novel possibility, through the first option, for creating room for emergence by a modification of dualities.

%Finally, regarding the question of emergence, I argued that in the models of space-time considered here, approximations are the key features that can make space-time emerge. In particular, there are two mutually opposed approaches to quantum gravity and I analyzed in which way gravity can emerge in each of them: (i) If the duality breaks down, gravity can emerge from a more fundamental boundary theory. This is the viewpoint in which there is no fundamental theory of quantum gravity, but only a fundamental quantum field theory from which gravity emerges. (ii) If a fundamental theory of quantum gravity does exist, then an approximation scheme (such as the Wilsonian coarse-graining approach) can make Einstein gravity to emerge from a more fundamental theory. Physics, not philosophy, is to decide which of the two variants is correct and applicable to our world. The task of philosophical analysis, as viewed here, is to map the possible alternatives and explicate how it is that gravity can emerge. The current analysis of emergence applies independently of the existence of dualities, hence it explicitly shows the consequences of having (or not having) a fundamental theory of quantum gravity for the possibility of gravity to emerge.

As a corollary, I provided a novel derivation of the Bekenstein-Hawking black hole entropy formula in Verlinde's scheme, including the correct numerical factor, which sheds light on how Verlinde's derivation compares to Bekenstein's calculation: Verlinde's is more general because it does not assume general relativity; and more robust because it gives the right numerical factor. The derivation also gave a new interpretation of the speed of light squared as the maximum amount of coarse-graining: an RG fixed point. This closes a circle of ideas in Verlinde's argument, which started from Bekenstein's thought experiment but left us in the dark as to: (a) how to recover the entropy of black holes; and (b) their relation to coarse-graining.

There are several other dualities in string theory and it is reasonable to expect that the present analysis can be applied to them, including the question of emergence. There are, however, many more {\it approximate} dualities, to which the framework for emergence should apply: for instance, cases of gauge/gravity duality with less symmetries; but also more realistic cases: cosmological models---gravity in spaces with a positive cosmological constant---where dualities have been conjectured to exist, but which are not very likely to exist in the strict sense---and so might be cases of broken duality. Hopefully, the conceptual framework for modifying dualities in order to make room for emergence, developed here, can be brought to bear in these more realistic cases as well.

\section*{Acknowledgements}
\addcontentsline{toc}{section}{Acknowledgements}

It is a pleasure to thank Jeremy Butterfield, Dennis Dieks, and Jeroen van Dongen. I also thank Elena Castellani and Dean Rickles for organising a wonderful workshop and all of the participants for the helpful discussions.

\section*{References}
\addcontentsline{toc}{section}{References}

Aharony, O.,  Gubser, S.S., Maldacena,  J.M., Ooguri, H., Oz, Y. (2000). ``Large \emph{N} field theories, string theory and gravity'',  \emph{Physics Reports}, 323(3-4), 183-386.  [hep-th/9905111].\\
  %%CITATION = HEP-TH/9905111;%%
\\
Ammon, M., Erdmenger, J.~(2015). ``Gauge/Gravity Duality. Foundations and Applications''. Cambridge University Press, Cambridge.\\
\\
Anderson, J.~(1964). ``Relativity principles and the role of coordinates in physics.'' In: H.Y. Chiu, W. Hoffman (eds.) {\it Gravitation and Relativity}, pp. 175–194. W.A. Benjamin, Inc., New York.\\
\\
Anderson, J.~(1967). ``Principles of Relativity Physics''. Academic Press, New York.\\
\\
 Avis, S.~J., Isham, C.~J., Storey, D.~(1978). ``Quantum Field Theory in anti-De Sitter Space-Time,'' {\it Physical Review Letters D}
{\bf 18} 3565.\\
  %%CITATION = PHRVA,D18,3565;%%
\\
Bekenstein, J. D. (1973). ``Black holes and entropy,'' {\it Physical Review D}, {\bf 7} 2333-2346. \\
\\
Belot, G.~(2011). ``Background-Independence,''  {\it General Relativity and Gravitation}  {\bf 43} 2865
  [arXiv:1106.0920 [gr-qc]].\\
  %%CITATION = ARXIV:1106.0920;%%
\\
Bouatta, N. and Butterfield, J. (2015), ``On emergence in gauge theories at the 't Hooft limit'', \emph{European Journal for Philosophy of Science}, 5(1), 55-87. [arXiv:1208.4986 [physics.hist-ph]].\\
  %%CITATION = ARXIV:1208.4986;%%
\\
Butterfield, J. (2011). ``Emergence, reduction and supervenience: a varied landscape'', \emph{Foundations of Physics}, 41(6), 920-959. \\
\\
Butterfield, J. and Bouatta, N. (2015). `On emergence in gauge theories at the {'t} Hooft limit', European Journal for Philosophy of Science 5, 2015, 55-87.\\
\\
Carlip, S.~(2014).   ``Challenges for Emergent Gravity,''
 {\it Studies in History and Philosophy of Modern Physics}, {\bf 46} 200
  [arXiv:1207.2504 [gr-qc]].\\
  %%CITATION = ARXIV:1207.2504;%%
\\
Castellani, E. (2010). ``Dualities and intertheoretic relations", pp. 9-19 in: Suarez, M., M. Dorato and M. Red\'{e}i (eds.). \emph{EPSA Philosophical Issues in the Sciences}. Dordrecht: Springer.\\
\\
Chandler, D. (1987), ``Introduction to Modern Statistical Mechanics'', New York, Oxford: Oxford University Press.\\
\\
Chirco, G., Liberati, S.~(2010). ``Non-equilibrium Thermodynamics of Spacetime: The Role of Gravitational Dissipation,''
{\it  Physical Review D}, {\bf 81} 024016  [arXiv:0909.4194 [gr-qc]].\\
  %%CITATION = ARXIV:0909.4194;%%
\\
Crowther, K. (2015). ``Decoupling emergence and reduction in physics'', {\it European Journal for Philosophy of Science}, 1-27.\\
\\
de Haro, S., Skenderis, K., and Solodukhin, S. (2001). ``Holographic reconstruction of spacetime and renormalization in the AdS/CFT correspondence", \emph{Communications in Mathematical Physics}, 217(3), 595-622. [hep-th/0002230].\\
\\
de Haro, S.~(2009). ``Dual Gravitons in AdS(4) / CFT(3) and the Holographic Cotton Tensor,''
  {\it Journal of High-Energy Physics}, {\bf 0901} 042
  [arXiv:0808.2054 [hep-th]].\\
  %%CITATION = ARXIV:0808.2054;%%
\\
de Haro, S., Petkou, A.~C.~(2014). ``Instantons and the Hartle-Hawking-Maldacena Proposal for dS/CFT,''
  {\it Journal of High-Energy Physics}, {\bf 1411} 126
  [arXiv:1406.6148 [hep-th]].\\
  %%CITATION = ARXIV:1406.6148;%%
\\
de Haro, S., Teh, N., Butterfield, J.N.~(2015). ``On the relation between dualities and gauge symmetries'', {\it Philosophy of Science}, submitted.\\
\\
Deser, S.~, Schwimmer, A.~(1993).  ``Geometric classification of conformal anomalies in arbitrary dimensions,''
  {\it Physics Letters B}, {\bf 309}  279
  [hep-th/9302047].\\
  %%CITATION = HEP-TH/9302047;%%
\\
Dieks, D., Dongen, J. van, Haro, S. de, (2015), ``Emergence in Holographic Scenarios for Gravity'', PhilSci 10606, arXiv:1501.04278 [hep-th]. {\it Studies in History and Philosophy of Modern Physics,} forthcoming.\\
  %%CITATION = ARXIV:1501.04278;%% 
\\
Di Francesco, P.~, Mathieu, P.~, S\'en\'echal (1996). ``Conformal Field Theory''. Springer-Verlag New York.\\
\\
Duncan, A.~(2012). ``The Conceptual Framework of Quantum Field Theory''. Oxford: Oxford University Press.\\
\\
Fraser, D.~(2015). Contribution to this volume.\\
\\
Giulini, D.~(2007). ``Some remarks on the notions of general covariance and background independence,''  {\it Lecture Notes in Physics}, {\bf 721} 105
  [gr-qc/0603087].\\
  %%CITATION = GR-QC/0603087;%%
\\
Heemskerk, I.~and Polchinski, J.~(2011). ``Holographic and Wilsonian Renormalization Groups,''
  {\it Journal of High-Energy Physics}, {\bf 1106} 031
  [arXiv:1010.1264 [hep-th]].\\
  %%CITATION = ARXIV:1010.1264;%%\\
\\
Henningson, M.~, Skenderis, K.~(1998). ``The Holographic Weyl anomaly,''  {\it Journal of High-Energy Physics}, {\bf 9807} 023
  [hep-th/9806087].\\
  %%CITATION = HEP-TH/9806087;%%
\\
Hugget, N.~(2015). Contribution to this volume.
\\
\\
Maldacena, J. (1998). ``The large \emph{N} limit of superconformal field theories and supergravity'',  
{\it International Journal of Theoretical Physics} 38 (1999) 1113; \emph{Advances in Theoretical and Mathematical Physics} (1998) 2, 231-252.
  [hep-th/9711200].\\
  %%CITATION = HEP-TH/9711200;%%
\\
Matsubara, K.~(2013). ``Realism, underdetermination and string theory dualities." {\it Synthese}, 190.3: 471-489.\\
\\
Pooley, O.~(2015). ``Background Independence, Diffeomorphism Invariance, and the Meaning of Coordinates'', in {\it Towards a Theory of Spacetime Theories}, Lehmkuhl (ed.), Einstein Studies Series (Boston: Birk\"auser), to appear. [arXiv:1506.03512].\\
\\
Rickles, D.~(2011). ``A philosopher looks at string dualities'',  \emph{Studies in History and Philosophy of Science Part B: Studies in History and Philosophy of Modern Physics}, 42(1), 54-67.\\
\\
Rickles, D.~(2012). ``AdS/CFT duality and the emergence of spacetime'', \emph{Studies in History and Philosophy of Science Part B: Studies in History and Philosophy of Modern Physics}, 44(3), 312-320.\\
\\
Smolin, L.~(2005). ``The Case for Background Independence,''  In *Rickles, D. (ed.) et al.: The structural foundations of quantum gravity* 196-239
  [hep-th/0507235].\\
  %%CITATION = HEP-TH/0507235;%%
\\
Teh, N.J.~(2013). ``Holography and emergence'', \emph{Studies in History and Philosophy of Science Part B: Studies in History and Philosophy of Modern Physics}, 44(3), 300-311.\\
\\
't Hooft, G.~(1993). ``Dimensional reduction in quantum gravity'', in: Ali, A., J. Ellis and S. Randjbar-Daemi, \emph{Salamfestschrift}. Singapore: World Scientific. [gr-qc/9310026].\\
  %%CITATION = GR-QC/9310026;%%\\
\\
't Hooft, G.~(2013). ``On the Foundations of Superstring Theory'', {\it Foundations of Physics}, 43 (1), 46-53 \\
\\
Verlinde, E. (2011). ``On the origin of gravity and the laws of Newton", {\it Journal of High Energy Physics,} 029. [arXiv:1001.0785 [hep-th]].

\end{document}